\documentclass[11pt,a4paper]{article}
\usepackage{amsfonts,amssymb,epsfig,amsmath,enumitem,mathtools}
\usepackage[utf8]{inputenc}
\usepackage{textcomp}
\usepackage{cite}
\usepackage{hyperref,caption,subcaption}
\usepackage{xcolor,tikz}

\renewcommand{\thefootnote}{\fnsymbol{footnote}}

\allowdisplaybreaks

\newcommand{\nn}[0]{\nonumber}

\begin{document}

\makeatletter \@addtoreset{equation}{section} \makeatother
\renewcommand{\theequation}{\thesection.\arabic{equation}}
\renewcommand{\thefootnote}{\fnsymbol{footnote}}	

\begin{titlepage}
\begin{center}

\hfill {\tt KIAS-P16047}\\
\hfill {\tt SNUTP16-003}\\

\vspace{2cm}

{\Large\bf M5-branes, orientifolds, and S-duality}

\vspace{2cm}

{\large Yoonseok Hwang$^1$, Joonho Kim$^2$, Seok Kim$^1$}

\vspace{0.7cm}

\textit{$^1$Department of Physics and Astronomy \& Center for
Theoretical Physics,\\
Seoul National University, Seoul 151-747, Korea.}\\

\vspace{0.2cm}

\textit{$^2$School of Physics, Korea Institute for Advanced Study,
Seoul 130-722, Korea.}\\

\vspace{0.7cm}

E-mails: {\tt yoonseok.hwang0@gmail.com, joonhokim@kias.re.kr, skim@phya.snu.ac.kr}

\end{center}

\vspace{1cm}

\begin{abstract}

We study the instanton partition functions of 5d maximal super Yang-Mills theories with all classical gauge groups.
They are computed from the ADHM quantum mechanics of the D0-D4-O4 systems.
Our partition functions respect  S-dualities of the circle compactified Yang-Mills theories and various orientifold backgrounds.
We also compute and study the $S^5$ partition functions that correspond to the 6d $(2,0)$ superconformal indices. 
Our $SO(2N)$ index takes the form of the vacuum character of $\mathcal{W}_D$ algebra in a special limit, supporting the $\mathcal{W}$ algebra conjecture. 
We propose new indices for $(2,0)$ theories with outer automorphism twists along the temporal circle, obtained from non-simply-laced SYMs on $S^5$.

\end{abstract}

\end{titlepage}

\setcounter{tocdepth}{2}
\tableofcontents

\section{Introduction} 

String theory provides a powerful framework for exploring quantum field theories. It leads to the discovery of quantum field theories in spacetime dimensions higher than 4. Important examples are six-dimensional maximally superconformal field theories, called $(2,0)$ theories, which come with the ADE classification \cite{Witten1995b}. The $(2,0)$ theories of AD-types describe the low energy dynamics of M5-branes \cite{Strominger1996}. It is very difficult to study these systems, due to the lack of their microscopic definitions.

Compactifying these 6d theories on a circle, one finds 5d maximal super Yang-Mills theories at low energy. Although these Yang-Mills theories are non-renormalizable, it has been suggested that they contain useful information about the 6d UV theories \cite{Douglas2011,Lambert2011}. Instanton solitons of 5d gauge theories play an essential role in understanding the 6d physics. They are non-perturbative solitons carrying the topological $U(1)$ charges which are interpreted as the Kaluza-Klein momenta along the circle. 

In this work, we study these 5d $\mathcal{N}=2$ gauge theories preserving 16 supercharges, obtained from circle compactifications of 6d $(2,0)$ SCFTs. A remarkable point is that even if one begins with 5d SYMs which do not recognize the 6d circle, the six-dimensional physics is recovered by incorporating non-perturbative instantons \cite{Douglas2011,Lambert2011,Kim2011,Kim2013b,Kim2012,Haghighat2015a,Kim2013a}. At least for AD-types, this claim is inspired by the duality relation between type IIA and M-theory. The Yang-Mills coupling constant $g_5^2$ is proportional to the radius of M-theory circle $R_\text{M}$ via the IIA string coupling constant. D0-branes bound to D4-branes are realized as instantons in 5d gauge theories, whose mass is inversely proportional to $R_\text{M}$, i.e.,
\begin{align}
	\label{eq:coupling-radius}
	\frac{4\pi^2}{g_5^2} \sim \frac{1}{R_\text{M}}.
\end{align}
This means that 5d SYM instantons carry Kaluza-Klein momenta along the M-theory circle. Including them, D4-branes are uplifted to circle compactified M5-branes described by 6d $(2,0)$ SCFT.

We study the instanton partition functions for 5d $\mathcal{N}=1^\ast$ gauge theories on Omega-deformed $\mathbb{R}^4 \times S^1$. The $\mathcal{N}=1^\ast$ theory is deformed from the maximal SYM by adding an $\mathcal{N}=1$ hypermultiplet mass $m$. We consider the Coulomb branch where the gauge symmetry is completely broken to its Abelian subgroup. The instanton partition function was first studied in \cite{Nekrasov2003, Nekrasov2007} to understand Seiberg-Witten solutions of 4d $\mathcal{N}=2$ gauge theories. Generalization to all classical gauge groups and inclusion of various hypermultiplets were considered in \cite{Nekrasov2004,Shadchin2005}. Throughout this work, we regard this observable as the Witten index of 5d SYM wrapped on the temporal circle $S^1$ of radius $\frac{\beta}{2\pi}$.

The 5d $SU(N)$ partition function was computed in \cite{Nekrasov2003,Nekrasov2007} via supersymmetric localization, being further interpreted as the Kaluza-Klein index of $SU(N)$-type $(2,0)$ theory \cite{Kim2011}. Besides instantons, the Nekrasov partition function also gets contribution from charged W-bosons which constitute \mbox{$\frac{1}{4}$-BPS} bound states with instantons. W-bosons are uplifted to self-dual strings of $(2,0)$ theories. They are electric and magnetic sources of tensor multiplets existing in 6d SCFTs. They are tensionless at the conformal fixed point. They obtain non-zero tension $T \propto \langle\Phi \rangle$ in the tensor branch, where tensor multiplet scalars $\Phi$ obtain non-zero VEVs $\langle\Phi \rangle \neq 0$ \cite{Witten1995b,Strominger1996,Witten1996b,Seiberg1996a}. The $SU(N)$ instanton partition function played important roles in the recent studies on $(2,0)$ theory of $A_{N-1}$-type \cite{Kim2011,Kim2013b,Kim2012,Haghighat2015a,Kim2013a}.

In this work, we extend the analyses made for the $SU(N)$ instanton partition functions to those of other classical gauge groups: $SO(2N+1)$, $Sp(N)$, $SO(2N)$. The $SO(2N)$ gauge theories are circle reductions of $(2,0)$ theories of $D_N$-type. The $SO(2N+1)$ and $Sp(N)$ gauge theories are obtained from circle compactified $(2,0)$ theories of AD-type with outer automorphism twists \cite{Witten2009,Tachikawa2011b}. We use these instanton partition functions to study the following subjects of the $(2,0)$ theories.

We first use the instanton partition functions to explore S-dualities of maximal super Yang-Mills theories. S-duality asserts that a pair of 4d $\mathcal{N}=4$ gauge theories are equal, where their gauge groups $G$ and $G^\vee$ are Langlands dual \cite{Goddard1977,Montonen1977,Witten1978,Osborn1979}. The W-bosons and monopoles in one theory correspond to the monopoles and W-bosons in the other theory, if their gauge couplings $\tau_4 = \frac{\theta_4}{2\pi} + \frac{4\pi i}{g_4^2}$ and $\tau_4^\vee = \frac{\theta_4^\vee}{2\pi} + \frac{4\pi i}{g_4^{\vee 2}}$ are related as $\tau_4^\vee \sim -\frac{1}{ \tau_4}$ \cite{Goddard1977,Montonen1977,Witten1978,Osborn1979}. It identifies a weakly-coupled theory and a strongly-coupled theory. Regarding 4d $\mathcal{N}=4$ SYMs as $(2,0)$ theories wrapped on tori, whose complex structures are translated to gauge couplings, S-duality is realized as exchanging two sides of the torus. Our instanton partition functions are 6d observables. Since they depend only on complex structures $\tau$ of the tori, they are also expected to respect the geometric S-dualities. In particular, for non-simply-laced gauge theories, S-duals of the instanton partition functions are expected to be those for 5d SYMs on $S^1$ with twisted boundary conditions \cite{Tachikawa2011b}. The instanton partition functions depend on various chemical potentials. Keeping all of these chemical potentials, their S-dualities are hard to explore. However, the instanton partition functions simplify after taking special limits of the chemical potentials. In Section~\ref{subsec:s-dual}, we discuss the S-dualities of the instanton partition functions in these limits.

As a byproduct, we study the S-dualities of type IIA orientifold backgrounds compactified on $S^1$ which uplift to M-theory on $T^2$. This is because one can study the D0-brane partition functions in various O4-plane backgrounds using the same techniques. Remarkably, one can perform the exact S-duality transformations on these partition functions with all chemical potentials turned on. See Section~\ref{sec:extra} for the details.

The instanton partition functions are also useful as building blocks for curved space partition functions. We compute the 5-sphere partition functions that were studied in \cite{Kallen2012,Kim2013b,Lockhart2012,Kim2012}. They are related to the partition functions of $(2,0)$ theories on $S^5 \times S^1$, which are 
%
%
%
called the $(2,0)$ superconformal indices \cite{Bhattacharya2008}. 
 For example, \cite{Kim2013b,Kim2012} obtained the $(2,0)$ superconformal index for $SU(N)$-type theory which agrees with the vacuum character of $\mathcal{W}_{A_{N-1}}$ algebra. It leads to the conjecture that there is an underlying $\mathcal{W}$ algebra structure in 6d $(2,0)$ SCFTs of A-type. This has been explicitly conjectured and further tested using 3-point functions in \cite{Beem2015a}. 
 We compute the $S^5$ partition functions for $SO(2N)$ gauge theories,  and show that they take the form of the vacuum character of $\mathcal{W}_{D_N}$ algebra\footnote{The correct $SO(2N)$ index was first reported in \cite{Kim2012}, which takes the form of $\mathcal{W}_{D_N}$ vacuum character. However, the derivation of the instanton part of the partition function was wrong in \cite{Kim2012}, which we correct in this paper.}.

The outline of this paper is as follows: In Section~\ref{sec:ADHM-qm}, we review the ADHM quantum mechanics of D0-D4-O4 systems and compute their Witten indices. In Section~\ref{sec:extra}, we study the S-dualities of pure orientifold systems compactified on a circle. In Section~\ref{sec:6dscft-5dsym}, we study S-dualities of the instanton partition functions in special limits. We also compute the $(2,0)$ superconformal indices from the $S^5$ partition functions, displaying the 6d operator spectra. Our results extend the $\mathcal{W}$ algebra conjecture to $SO(2N)$ theories and also propose the new indices for $(2,0)$ theories with outer automorphism twists. Concluding remarks are given in Section~\ref{sec:conclusion}.

\section{Instantons in 5d maximal SYM} 

\label{sec:ADHM-qm}

We consider 5d maximal SYM on $\mathbb{R}^{1,4}$ having a classical gauge group $G$ with rank $N$. It has $SO(1,4)$ Lorentz symmetry and $SO(5)_R$ R-symmetry. We study the Coulomb branch where the vector multiplet scalars $\phi_I$ acquire non-zero VEVs $\alpha_I$, breaking the gauge symmetry $G$ into the Abelian subgroup $U(1)^N \subset G$. The maximal SUSY algebra is given by
\begin{align}
	\{Q_M^i, Q_N^j\} = P_\mu (\Gamma^\mu C)_{MN}\omega^{ij} + i \frac{4\pi^2 k}{g_5^2} C_{MN}\omega^{ij} + i\, \text{Tr}\, {(\alpha_I \cdot \Pi)} C_{MN} (\Gamma^I \omega)^{ij}
\end{align}
where $M,N = 1,2,3,4$ are $SO(1,4)$ spinor indices, $i,j = 1,2,3,4$ are $SO(5)_R$ spinor indices, $I$ is the $SO(5)_R$ vector index, $C_{MN}$ is the charge conjugation matrix, $\omega^{ij}$ is the $SO(5)_R \simeq Sp(2)_R$ symplectic form, $\Pi$ denote $U(1)^N$ gauge symmetry generators. Supercharges are subject to the symplectic-Majorana condition. The $U(1)$ instanton charge $k$ is defined as
\begin{align}
	\label{eq:instanton-charge}
	k = \frac{1}{8\pi^2} \int_{\mathbb{R}^4} \text{Tr}\, (F\wedge F) \in \mathbf{Z}. 
\end{align}
which is integer-valued. We write Yang-Mills kinetic term as $\frac{1}{4g_5^2} \int \text{Tr} \, (F^{\mu\nu} F_{\mu\nu})$, setting the unit instanton mass to be $4\pi^2/g_5^2$.

The maximal SYM contains an $\mathcal{N}=2$ vector multiplet whose $\mathcal{N}=1$ decomposition gives a vector multiplet plus an adjoint hypermultiplet. We look at the $\mathcal{N}=1$ Coulomb branch where only the $\mathcal{N}=1$ vector multiplet scalar has a non-zero VEV $\alpha$. This set-up has $SO(4)_1$ little group of 5d massive particles and $SO(4)_{2} \subset SO(5)_R$ R-symmetry. We further decompose it into $SU(2)_{1L} \times SU(2)_{1R} \subset SO(4)_1$ and $SU(2)_{2L} \times SU(2)_{2R} \subset SO(4)_2$.  We denote 16 generators of maximal SUSY by $Q^A_{\alpha}, Q^A_{\dot{\alpha}}, Q^a_{\alpha}, Q^a_{\dot{\alpha}}$, where $\alpha, \dot{\alpha}, a, A$ are doublet indices for $SU(2)_{1L}$, $SU(2)_{1R}$, $SU(2)_{2L}$, $SU(2)_{2R}$.

There are two types of massive $\frac{1}{2}$-BPS particles in the Coulomb phase: W-bosons and instantons. W-bosons are electrically charged objects under $U(1)^N \subset G$, which we choose to satisfy $\text{Tr}(\alpha \cdot \Pi) > 0$ and preserve $Q^A_{\dot{\alpha}}$ and $Q^a_{\alpha}$. Instantons are solitonic particles that carry the topological $U(1)$ charge defined in \eqref{eq:instanton-charge}, which we choose to satisfy the self-duality condition $F_{mn} = \frac{1}{2}\epsilon_{mnpq} F_{pq}$ and preserve $Q^A_{\dot{\alpha}}$ and $Q^a_{\dot{\alpha}}$. Notice that our choice restricts the instanton charge $k$ to be a positive integer. These BPS particles may form $\frac{1}{4}$-BPS bound states, whose masses are given by
\begin{align}
	\label{eq:mass-bps}
	M = \frac{4\pi^2k}{g_5^2} + \text{Tr}\, (\alpha \cdot \Pi).
\end{align}
In this section, we study the instanton partition functions which count these BPS bound states of instantons and W-bosons in 5d maximal super Yang-Mills theories.

\subsection{$\mathcal{N}=(4,4)$ ADHM quantum mechanics}
\label{subsec:adhmqm}

The moduli space approximation is a technique to describe the low energy dynamics of solitons \cite{Manton1982}. Being applied to 5d SYM instantons, it gives a SUSY quantum mechanics for instanton zero modes. When the 5d gauge group is classical, the zero modes are described as the ADHM data satisfying the ADHM constraint equation \cite{Atiyah1978a}. The instanton quantum mechanics is a non-linear sigma model whose target space is the instanton moduli space. 

However, one cannot expect the instanton quantum mechanics to be a UV complete description for instanton solitons since the instanton moduli space suffers from the small instanton singularity where the instanton size shrinks to zero. It is generally very demanding task to find a UV completion of the instanton quantum mechanics. In certain type of theories, string theory supplies the UV completions from D-brane realizations of 5d SYM instantons. We call them the ADHM quantum mechanics.

5d maximal SYM with a classical gauge group $G$ is engineered from  D4-branes possibly on top of an O4-plane. Its instantons are realized as D0-branes stuck on D4-branes. The worldvolume theory of D0-branes is the ADHM quantum mechanics that we study. It is a gauge theory whose gauge group $\hat{G}$ is determined from $G$. It has $SO(4)_1 \times SO(5)_R$ symmetries inherited from underlying 5d theories. It  preserves $\mathcal{N}=(4,4)$ supersymmetry\footnotemark \ generated by $Q^A_{\dot{\alpha}}$ and $Q^a_{\dot{\alpha}}$. Its field contents are induced from massless modes of open strings ending on D0-branes. Here we list them as $(4,4)$ SUSY multiplets.
\begin{align}
\begin{tabular}{c   l }
	\label{eq:adhmqm-matter}
	$(A_0, \varphi, \varphi_{aA} \,| \,  \lambda^{A}_{\dot{\alpha}}, \lambda^{a}_{\dot\alpha})$ & \quad Vector multiplet in \textbf{adj} ($\hat{G}$)\\
	$(a_{\alpha\dot{\alpha}} \,| \, \lambda^{A}_{\alpha}, \lambda^{a}_{\alpha})$ & \quad Hypermultiplet in \textbf{R} ($\hat{G}$) \\
	$(q_{\dot{\alpha}}\,|\, \psi^A, \psi^a)$ & \quad Hypermultiplet in \textbf{bif}
	 ($\hat{G} \times G$) 
\end{tabular}
\end{align}
Here $\textbf{bif} (\hat{G} \times G) $ denotes the bifundamental representation of $\hat{G} \times G$. Type of the O4-plane determines the 5d group $G$, the 1d group $\hat{G}$, and the representation $\textbf{R}$ as follows \cite{Dorey2002}.
\begin{align}
\label{tab:typeof-adhmqm}
\begin{tabular}{ c | c | c | c }
Type of O4 & 5d gauge group $G$ & 1d gauge group $\hat{G}$ & Representation $\mathbf{R}$ of $\hat{G}$ \\\hline
----- & $U(N)$ & $U(k)$ & adjoint\\
$\text{O4}^-$ & $SO(2N)$ & $Sp(k)$ & antisymmetric\\
$\text{O4}^0$ & $SO(2N+1)$ & $Sp(k)$ & antisymmetric\\
$\text{O4}^+$ & $Sp(N)_{\theta = 0}$ & $O(k)_{\theta = 0}$ & symmetric\\
$\widetilde{\text{O4}}^+$ & $Sp(N)_{\theta = \pi}$ & $O(k)_{\theta = \pi} $ & symmetric
\end{tabular}
\end{align}
$k$ denotes The instanton number. The action of the $U(k)$ ADHM quantum mechanics is given in \cite{Kim2011}. Generalization to other gauge groups is straightforward.
\footnotetext{1d $\mathcal{N}=(4,4)$ SUSY should be understood as the circle reduction of 2d $(4,4)$ SUSY with $SO(4) \times SO(4)$ R-symmetry.}

The ADHM quantum mechanics include more fields than the instanton quantum mechanics, which are massless only at the small instanton singularity. The vector multiplet scalars $\varphi$, $\varphi_{aA}$ are such extra fields, which open up a new branch of the moduli space touching the instanton moduli space exactly at the small instanton singularity. We call it the Coulomb branch of the ADHM quantum mechanics. This UV-completion of the instanton moduli space allows D0-branes to move away from D4-branes. One can regard D0-branes as 5d SYM instantons only in the Higgs branch of the quantum mechanics, which is spanned by the ADHM data $a_{\alpha\dot{\alpha}}$, $q_{\dot{\alpha}}$ subject to the ADHM constraint equation. 

We want to use the ADHM quantum mechanics for counting 5d SYM states involving instantons. However, the Coulomb branch may give extra contributions to the Witten indices which are irrelevant to the 5d SYM partition functions. See Section~\ref{sec:extra} for related discussions.

\subsection{Witten index}
\label{subsec:witten-index}

%
%

In this section, we compute the index of the ADHM quantum mechanics defined as
\begin{align}
	\label{eq:def-witten-index}
    I_k = \text{Tr}_k\Big[(-1)^F \, e^{-\beta \{ Q , Q^{\dagger} \}} \, t^{2 (J_{1R} + J_{2R})} u^{2 J_{1L}} v^{2J_{2L}} \prod_{a=1}^n w_a^{2\Pi_a}\Big].
\end{align}
It counts BPS states annihilated by $Q\equiv {Q}^{A=1}_{\dot{\alpha}=\dot{1}}$ and $Q^\dagger \equiv {Q}^{A=2}_{\dot{\alpha}=\dot{2}}$. $J_{1L}$, $J_{1R}$, $J_{2L}$, $J_{2R}$ are the Cartan generators for the $SU(2)_{1L}$, $SU(2)_{1R}$, $SU(2)_{2L}$, $ SU(2)_{2R}$ global symmetries. $\Pi_a$ denote the Cartan generators of the 5d gauge group $G$. Various fugacities are conjugate to the Cartans of global symmetries commuting with $Q$, $Q^\dagger$. Besides $\Pi_a$'s, $J_{1R} + J_{2R}$, $J_{1L}$, $J_{2L}$ are all commuting combinations. We often express the fugacities using chemical potentials as follows:
\begin{align}
t=e^{-\epsilon_{+}},\ u=e^{-\epsilon_{-}},\ v=e^{- m},\ w_{i}=e^{- \alpha_{i}}.
\end{align}

The chemical potentials deform the underlying 5d gauge theory. $\epsilon_1 = \frac{\epsilon_+ + \epsilon_-}{2}$ and $\epsilon_2 = \frac{\epsilon_+ - \epsilon_-}{2}$ put the 5d gauge theory on Omega-deformed background. $m$ becomes the mass of the $\mathcal{N}=1$ adjoint hypermultiplet, yielding the mass-deformed $\mathcal{N}=1^*$ SYM. $\alpha$ are complexified chemical potentials of $G$, breaking the 5d gauge group $G$ into its Abelian subgroup $U(1)^N$. Such deformations regulate the 5d gauge theory at long distances, defining the instanton partition function with IR regulators \cite{Nekrasov2003}. They give mass to the scalars $\varphi_{aA}$, $a_{\alpha\dot{\alpha}}$, $q_{\dot{\alpha}}$ in the ADHM quantum mechanics, making the BPS spectrum to be gapped. However, there still remains a flat direction in the moduli space. As the vector multiplet scalar $\varphi$ is not charged under any global symmetries, chemical potentials cannot give it a mass. If an Fayet-Iliopoulos deformation is available, i.e., if $\hat{G}$ contains an overall $U(1)$ factor, $\varphi$ acquires mass. In general, we encounter a continuous BPS spectrum that makes the index \eqref{eq:def-witten-index} hard to compute.

Nevertheless we can obtain the instanton partition function from the indices \eqref{eq:def-witten-index} of the quantum mechanics. The flat direction spanned by $\varphi$ belongs to the Coulomb branch that decouples from the Higgs branch in IR \cite{Aharony1997,Aharony1998}. Index contributions of the decoupled Coulomb branch can be identified and removed from the multi-particle indices, which are factorized as \cite{Hwang2015a}
\begin{align}
	\label{eq:def-multiparticle-index}
	Z_\text{ADHM} = 1 + \sum_{k=1}^\infty q^k\, I_k = Z_\text{inst} \cdot Z_\text{extra},
\end{align}
where $q \equiv e^{-{4\pi^2 \beta}/{g_5^2}}$ denotes the instanton number fugacity. The instanton partition function $Z_\text{inst}$ comes from the Higgs branch contributions which do not involve any subtleties from the Coulomb branch continuum. 

We now turn to the path integral localization of the index \eqref{eq:def-witten-index}, following \cite{Hwang2015a,Hori2015}. The path integral measure is given by the Euclidean action of the ADHM quantum mechanics. The temporal direction is compactified as the circle with circumference $\beta$, used in \eqref{eq:def-witten-index}. Once we take the weak coupling limit $g_1 \rightarrow 0$, where $g_1$ denotes the coupling constant of the quantum mechanics, the path integrals are reduced to Gaussian integrals around zero modes. The most important zero modes are holonomies of the gauge field $A_0$ along the temporal circle and the vector multiplet scalar $\varphi$. They are combined into dimensionless, complexified holonomies $\phi = i \beta A_0 + \beta \varphi $. $A_0$ is subject to a large gauge transformation on the temporal circle, making the imaginary parts of $\phi$ eigenvalues to be periodic. The complexified Wilson lines $e^{\phi} \in \hat{G}$ are gauge invariant. We label $\phi$ as follows \cite{Shadchin2005,Kim2012c}:
\begin{align}
	\label{eq:holonomy-for-each-group}
	U(\ell) \ \ni \quad & e^{\phi} = \text{diag} (e^{\phi_1}, \cdots, e^{\phi_\ell}) && \longrightarrow && \phi = (+\phi_1, \cdots, +\phi_\ell)\\
	O(2\ell)_+ \ \ni \quad  & e^{\phi} = \text{diag} (e^{\sigma_2 \phi_1}, \cdots, e^{\sigma_2 \phi_\ell}) && \longrightarrow && \phi =  (\pm\phi_1, \cdots, \pm\phi_\ell) \nn\\ 
	O(2\ell)_- \ \ni \quad & e^{\phi} = \text{diag} (e^{\sigma_2 \phi_1}, \cdots, e^{\sigma_2 \phi_{\ell-1}}, \sigma_3) && \longrightarrow && \phi =  (\pm\phi_1, \cdots, \pm\phi_{\ell-1}, 0, i\pi) \nn\\ 
	O(2\ell+1)_+ \ \ni \quad & e^{\phi} = \text{diag} (e^{\sigma_2 \phi_1}, \cdots, e^{\sigma_2 \phi_\ell},+1)  && \longrightarrow && \phi = (\pm\phi_1, \cdots, \pm\phi_{\ell}, 0) \nn\\
	O(2\ell+1)_- \ \ni \quad & e^{\phi} = \text{diag} (e^{\sigma_2 \phi_1}, \cdots, e^{\sigma_2 \phi_\ell},-1)  && \longrightarrow && \phi =  (\pm\phi_1, \cdots, \pm\phi_{\ell}, i\pi) \nn \\
	Sp(\ell)_- \ \ni \quad & e^{\phi} = \text{diag} (e^{\sigma_3 \phi_1}, \cdots, e^{\sigma_3  \phi_\ell})  && \longrightarrow && \phi =  (\pm\phi_1, \cdots, \pm\phi_{\ell}) \nn
\end{align}
There are also other zero modes coming from the gaugino and the auxiliary scalar field $D$ \cite{Hwang2015a,Hori2015}.

We perform the Gaussian integrals over massive fluctuations, fixing the zero modes for a while. Integration gives the 1-loop determinants $I_\text{1-loop}$ which are products of the following factors \cite{Shadchin2005}.
\begin{align}
	\label{eq:1loop-adj-vector}
	I_{\rm vector} &=   \prod_{\hat{\rho}\, \in\, \text{\textbf{root}}\,(\hat{G})}\bigg( 2\sinh{\tfrac{\hat{\rho}(\phi)}{2}}\bigg) \cdot \prod_{\hat{\rho}\, \in\, \text{\textbf{adj}}\,(\hat{G})} \Bigg(\frac{ 2\sinh{\frac{\hat{\rho}(\phi) + 2\epsilon_+}{2}}}{2\sinh{\frac{\hat{\rho}(\phi) - \epsilon_+ \pm m}{2}} }\Bigg) \cdot \prod_{i=1}^r d\phi_i \\
	\label{eq:1loop-R-hyper}
	I_{\rm hyper}^{\mathbf{R}} &=  \prod_{\hat{\rho}\, \in\, \text{\textbf{R}}\,(\hat{G})} \Bigg(\frac{2\sinh{\frac{\hat{\rho}(\phi) + m \pm \epsilon_-}{2}}}{2\sinh{\frac{\hat{\rho}(\phi) + \epsilon_+ \pm \epsilon_- }{2}} }\Bigg)\\
	\label{eq:1loop-bif-hyper}
	I_{\rm hyper}^{\mathbf{bif}} &= 
		\begin{dcases*}
			\prod_{\hat{\rho}\, \in\, \text{\textbf{fnd}}\,(\hat{G})} \prod_{\rho\, \in\, \text{\textbf{fnd}}\,(G)} \Bigg(\frac{2\sinh{\frac{\pm (\hat{\rho}(\phi) - \rho(\alpha)) + m}{2}}}{2\sinh{\frac{\pm (\hat{\rho}(\phi) - \rho(\alpha)) + \epsilon_+ }{2}} }\Bigg) & for $G = U(N)$ and $\hat{G} = U(k)$\\
			\prod_{\hat{\rho}\, \in\, \text{\textbf{fnd}}\,(\hat{G})} \prod_{\rho\, \in\, \text{\textbf{fnd}}\,(G)} \Bigg(\frac{2\sinh{\frac{\hat{\rho}(\phi) - \rho(\alpha) + m}{2}}}{2\sinh{\frac{\hat{\rho}(\phi) - \rho(\alpha) + \epsilon_+ }{2}} }\Bigg) & for all other cases
		\end{dcases*}
\end{align}
We use the $\pm$ notation: $2\sinh{(a \pm b)} \equiv 2\sinh{(a+b)}\cdot 2\sinh{(a-b)}$. $r$ is the number of continuous parameters in the $\phi$ holonomies. One-loop determinants for fields in certain representations of $\hat{G}$ and $G$ involve the parameters $\hat{\rho}(\phi)$ and $\rho(\alpha)$. We refer to \cite{Shadchin2005} for the parameters in all rank-1 and rank-2 representations of classical groups. For example, the parameters for the $U(k)$ representations are 
\begin{align}
	\mathbf{fund}: &\quad \hat{\rho}(\phi) = (+\phi_i)_{1\leq i\leq k}\nn \\
	\mathbf{adj}: &\quad \hat{\rho}(\phi) = (+\phi_i - \phi_j)_{1\leq i, j \leq k}\\
	\mathbf{symm}: &\quad \hat{\rho}(\phi) = (+\phi_i + \phi_j)_{1\leq i \leq j \leq k}\nn\\
	\mathbf{anti}: &\quad \hat{\rho}(\phi) = (+\phi_i + \phi_j)_{1\leq i < j \leq k}\nn
\end{align}
The \textbf{root} is defined such that the coupled parameters $\hat{\rho}(\phi)$ are 
\begin{align}
	\hat{\rho}(\phi) = \begin{dcases*} 
	(+\phi_i - \phi_j)_{1\leq i \neq j \leq k} & for $\hat{G} = U(k)$\\
 \hat{\rho}(\phi) \text{ of  \textbf{adj}}	& for all other cases
 \end{dcases*}
\end{align}
Note that 1-loop determinants of real scalars and fermions involve square roots of sinh factors. A pair of such factors are always arranged as $\sqrt{\sinh{\frac{z+a}{2} \sinh{\frac{-z-a}{2}} }} \sim \sinh{(\frac{z+a}{2})}$, following \cite{Hwang2015a}. This is why the first line of \eqref{eq:1loop-bif-hyper} has twice of sinh factors than the second line of \eqref{eq:1loop-bif-hyper}.

%

The next step is to integrate over the zero modes. Carefully treating the zero modes of the gaugino and the auxiliary scalar, the integral becomes a contour integral over the space of $\phi$ holonomies, i.e., $r$ copies of a cylinder \cite{Hwang2015a,Hori2015}. The integrand $I_\text{1-loop}$ develops various poles, some being inside a finite region and the others being at infinities. We need a proper choice of contour to complete the integral\footnote{For $\mathcal{N}=1^*$ $SU(N)$ theories, Nekrasov guessed the correct contour prescription \cite{Nekrasov2003}. More complicated contour prescriptions for other gauge groups are given in \cite{Fucito2004}. These are compatible with the rules that we explain.}.
 For $r=1$, the index contribution of the residues $R_{\pm\infty}$ at infinities appears as the sum $R_{+\infty} + R_{-\infty}$ with a coefficient depending on a Fayet-Iliopoulos parameter $\zeta$ \cite{Hwang2015a,Hori2015}. We expect the same for $r>1$. One can show that the residue sums $R_{+\infty} + R_{-\infty}$, for the $\mathcal{N}=(4,4)$ ADHM quantum mechanics, are zero for all cylinders. So the index is independent of $\zeta$. It receives the contributions only from the residues inside the finite region, chosen by the Jeffrey-Kirwan residue operation \cite{Hwang2015a,Cordova2014,Hori2015}. For non-degenerate poles, where $r$ distinct sinh factors in the denominator of $I_\text{1-loop}$ become zero, the Jeffrey-Kirwan residue can be expressed as
\begin{align}
\label{eq:JK-Res}
  \textrm{JK-Res}_{\phi_\ast}({\bf Q}_\ast,\eta)
  \frac{d\phi_1\wedge \cdots\wedge d\phi_r}{Q_{j_1}(\phi\!-\!\phi_\ast)\cdots Q_{j_r}(\phi\!-\!\phi_\ast)}
  =\left\{\begin{array}{ll} | \det(Q_{j_1},\cdots,Q_{j_r}) |^{-1}&
  \textrm{if }\eta\in{\rm Cone}(Q_{j_1},\cdots, Q_{j_r})\\
  0&{\rm otherwise}\end{array}\right..
\end{align}
It is a linear functional which refers to a pole location $\phi_\ast$ and an auxiliary vector $\eta$ in an $r$-dimensional charge space. ${\bf Q}_\ast=(Q_1,\cdots,Q_r)$ is a set of $r$ charge vectors, associated to the sinh factors in the denominator being zero at $\phi_\ast$. The vector $\eta$ has to be generic for JK-Res$_{\phi_\ast}({\bf Q}_\ast,\eta)$ to be well-defined. We refer to \cite{Hori2015} for treatment of degenerate poles. 

We recall that $\hat{G} = O(k)$ allows the two disconnected Wilson lines in $O(k)_+$ and $O(k)_-$. In general, if $\hat{G}$ is disconnected, there can exist multiple, disconnected Wilson line backgrounds of $\hat{G}$. The Witten index includes a sum over distinct holonomy sectors such that
\begin{align}
	\label{eq:index-final-formula}
	I_k = \sum_a \frac{1}{|W_a|}\oint I_\text{1-loop}^{(a)} =
  \sum_a \frac{1}{|W_a|} \sum_{\phi_\ast}\textrm{JK-Res}_{\phi_\ast}({\bf Q}_\ast,\eta)\,
  I_\text{1-loop}^{(a)},
\end{align}
where $a$ labels the disconnected holonomy sectors, $|W_a|$ and $I_\text{1-loop}^{(a)}$ are the Weyl group order and the 1-loop determinant in a given holonomy sector $a$. $\phi_\ast$ runs over all existing poles in the integrands. The Weyl group orders $|W_a|$ that preserve the holonomies given in \eqref{eq:holonomy-for-each-group} are 
\begin{align}
	&|W|_{U(\ell)} = \ell!, &
	&|W_+|_{O(2\ell)} = 2^{\ell-1} \ell!, &
	&|W_-|_{O(2\ell)} = 2^{\ell-1} (\ell-1)!,\\
	&|W_+|_{O(2\ell+1)} = 2^{\ell} \ell!, &
	&|W_-|_{O(2\ell+1)} = 2^{\ell} \ell!, &
	&|W|_{Sp(\ell)} = \ell! \nn.
\end{align}
Note that 5d $Sp(N)$ SYMs have two discrete choices of $\theta$ angles $0$, $\pi$ related to $\pi_4 (Sp(N)) = \mathbb{Z}_2$ \cite{Douglas1996}. The 5d $\theta$ parameters induce the 1d discrete $\theta$ parameters related to $\pi_0 (O(k)) = \mathbb{Z}_2$, which are the Wilson lines along the temporal circle. The sums over holonomy sectors in \eqref{eq:index-final-formula} are given as \cite{Bergman2014b}
\begin{align}
	I_k = \begin{dcases}
		\tfrac{1}{2}(I_k^+ + I_k^-) & \text{for }+1 \in \pi_0 (O(k))\\
		\tfrac{(-1)^k}{2}(I_k^+ - I_k^-) & \text{for }-1 \in \pi_0 (O(k)).
	\end{dcases}
\end{align}

\section{Orientifolds from M-theory}
\label{sec:extra} 

The multi-particle index $Z_\text{ADHM}$ of the ADHM quantum mechanics is computable using the formula \eqref{eq:index-final-formula}. $Z_\text{ADHM}$ captures BPS states of D0-branes, some of which are decoupled states from 5d SYM. The multi-particle index $Z_\text{ADHM}$ is factorized into two parts \cite{Hwang2015a}:
\begin{align*}
	Z_\text{ADHM} = 1 + \sum_{k=1}^\infty q^k\, I_k = Z_\text{inst} \cdot Z_\text{extra}.
\end{align*}
$Z_\text{inst}$ is the instanton partition function. $Z_\text{extra}$ is the index for the extra states. To obtain the instanton partition function, we need to identify and remove $Z_\text{extra}$ from $Z_\text{ADHM}$. In this section, we determine $Z_\text{extra}$ for various D0-D4-O4 systems on a case-by-case basis.

We already discussed that the extra contribution comes from the Coulomb branch, which is parametrized by $\varphi$ and $\varphi_{aA}$. At a generic point of the Coulomb branch, these scalar fields acquire non-zero values which represent D0-branes' transverse positions to D4-branes. We expect to determine $Z_\text{extra}$ by analyzing BPS states of D0-branes far away from D4-branes. For this purpose, we separately compute the multi-particle indices of D0-branes in the pure orientifold backgrounds without any dynamical D4-branes.

The pure orientifold backgrounds are obtained by formally taking $N=0$ in the ADHM quantum mechanics. The extra indices $Z_\text{extra}$ are included in the $N=0$ indices which can be written in a simple way using the plethystic exponential
\begin{align}
	\text{PE}[f(q, t,u,v,w_a)] \equiv \exp{\left[\sum_{n=1}^\infty \frac{1}{n}f(q^n,t^n,u^n,v^n,w_a^n)\right]}.
\end{align}
$\text{PE}[f]$ is the multi-particle index of a single-particle index $f$ \cite{Benvenuti2007}. The $N=0$ indices are given by
\begin{align}
\label{eq:decoupled-u2n}
&Z_\text{ADHM}^\text{U(0)} &&\hspace{-0.3cm}= 1\\
\label{eq:decoupled-so2n}
&Z_\text{ADHM}^\text{SO(0)}  &&\hspace{-0.3cm}= \text{PE}\left[\frac{t^2 (v+v^{-1}-u-u^{-1})(t+t^{-1})}{2(1-tu)(1-tu^{-1})(1+tv)(1+tv^{-1})}\frac{q}{1-q}\right]\\
\label{eq:decoupled-so2nodd}
&Z_\text{ADHM}^\text{SO(1)} &&\hspace{-0.3cm}=  \text{PE}\left[\frac{t^2 (v+v^{-1}-u-u^{-1})(t+t^{-1})}{2(1-tu)(1-tu^{-1})(1+tv)(1+tv^{-1})}\frac{-q}{1-(-q)}\right]\\
\label{eq:decoupled-spn-notheta}
&Z_\text{ADHM}^\text{Sp(0)${}_{\theta = 0}$} &&\hspace{-0.3cm}=  \text{PE}\left[\frac{t^2 (v+v^{-1}-u-u^{-1})(t+t^{-1})}{2(1-tu^\pm)(1+tv^\pm)}\frac{q^2}{1-q^2} + \frac{t (v+v^{-1}-u-u^{-1})}{(1-t u)(1-t u^{-1})}\frac{q}{1-q^2}\right]\\
\label{eq:decoupled-spn-theta}
&Z_\text{ADHM}^\text{Sp(0)${}_{\theta = \pi}$} &&\hspace{-0.3cm}=  \text{PE}\left[\frac{t^2 (v+v^{-1}-u-u^{-1})(t+t^{-1})}{2(1-tu^\pm)(1+tv^\pm)}\frac{-q^2}{1-(-q^2)} + \frac{t (v+v^{-1}-u-u^{-1})}{(1-t u)(1-t u^{-1})}\frac{q^2}{1-q^4} \right]
\end{align}
where we use the $\pm$ notation: $(1-xy^\pm) \equiv (1-xy)(1-xy^{-1})$. Their $q$-dependences were checked up to $q^3$-order in \eqref{eq:decoupled-so2n} and \eqref{eq:decoupled-so2nodd} (for SO-type) and $q^7$-order in \eqref{eq:decoupled-spn-notheta} and \eqref{eq:decoupled-spn-theta} (for Sp-type). \eqref{eq:decoupled-u2n} can be checked in all $q$-orders.

%
%
%
%


The subtlety arising here is that there exists a flat direction, spanned by $\varphi$, even after turning on all chemical potentials. This damages the robustness of the Witten index, which is the critical assumption for localizing the path integrals. The Witten index is defined in the regime $g_1^2 \beta^3 \gg 1$, while the index formula \eqref{eq:index-final-formula} is derived in the regime $g_1^2 \beta^3 \ll 1$. If there were no flat direction, such deformation would not change the index. However, the ADHM quantum mechanics has the flat direction developed by $\varphi$, so that the $N=0$ indices exhibit fractional coefficients $\frac{1}{2}$'s in \eqref{eq:decoupled-so2n}-\eqref{eq:decoupled-spn-theta}, as discussed in \cite{Yi1997,Sethi1998,Lee2016}. Due to the decoupling of the Higgs and Coulomb branch at low energy \cite{Aharony1997,Aharony1998}, we still expect that the continuum from $\varphi$ spoils $Z_{\rm extra}$, not $Z_{\rm inst}$.

In order to identify the extra indices $Z_\text{extra}$ from the $N=0$ indices, we study the origins of D0-brane bound states in the $N=0$ indices from M-theory perspectives. It also tells us interesting information on the orientifold backgrounds in string theory. While $SO(9)$ is the little group for 10d massive particles, we only try to understand the $SO(8) \subset SO(9)$ spin contents of D0-branes, setting aside $\frac{1}{2}$'s in \eqref{eq:decoupled-so2n}-\eqref{eq:decoupled-spn-theta}, where the fugacities $t$, $u$, $v$ are conjugate to the $SO(8)$ spins.

First of all, we remark on the geometric interpretations of the instanton number fugacities $q$. They are defined as $q \equiv e^{-S_0}$ using the unit instanton action $S_0 = \frac{4\pi^2 \beta}{g_5^2}$. Since D0-branes carry the Kaluza-Klein momenta along the M-circle, their masses are related to the inverse of the M-circle radius. 
The $O(k)$ theories describe the dynamics of $k$ `half' D0-branes, while the $U(k)$ and $Sp(k)$ theories describe the dynamics of $k$ `full' D0-branes. This observation implies that the relations \eqref{eq:coupling-radius} between the 5d gauge couplings and the M-circle radii should be precisely stated as
\begin{align}
	\label{eq:coupling-radius-precise}
	\frac{4\pi^2}{g_5^2} = \frac{1}{2R_\text{M}}   \text{\ \ for $\hat{G} = O(k)$} && \frac{4\pi^2}{g_5^2} = \frac{1}{R_\text{M}} \text{\ \ otherwise}.
\end{align}
Accordingly, the instanton number fugacities 
\begin{align}
	q \equiv \exp{(-\tfrac{4\pi^2\beta }{g_5^2})} =
	\begin{dcases*}
	\exp{(-\tfrac{\beta}{2R_M})} \equiv e^{ \pi i \tau} & \quad $\hat{G} = O(k)$\\
	\exp{(-\tfrac{\beta}{R_M})} \equiv e^{ 2\pi i \tau} & \quad otherwise
 	\end{dcases*}
\end{align}
are interpreted as the M-circle momentum fugacities with a half-integer unit (for $\hat{G} = O(k)$) and with integer units (for all other cases). In all cases, M-theory wraps the torus $T^2$. It is made of the M-circle and the temporal circle, whose radii are $R_M$ and $\frac{\beta}{2\pi}$ respectively. Its complex structure $\tau$ is given by $\tau \equiv i \, \frac{\beta/2\pi}{R_{M}}$. Note that S-dualities of IIA backgrounds, often called the ``9-11 flip'' \cite{Dijkgraaf1997}, correspond to exchanging the two sides of the torus $T^2$, i.e., $\tau \longrightarrow -\frac{1}{\tau}$.

Now we analyze \eqref{eq:decoupled-so2n}-\eqref{eq:decoupled-spn-theta} from M-theory perspectives, apart from the $\frac{1}{2}$ factors.

\vspace{-0.5\baselineskip}
\paragraph{\underline{Pure D0}} We consider M-theory on $\mathbb{R}^{1,9} \times S^1$ and its  massless particles with non-zero momentum $P_{11}$ along the M-circle, satisfying
 $|P_0| = |P_{11}| >0 $ and $P_{i=1,\cdots,9} = 0$. The 11d SUSY algebra is written as $\{Q, Q^\dagger\} \sim (-1 \pm \Gamma^0 \Gamma^{11}) |P^0|$. The 11d gravity multiplet preserves half of the 32 supercharges, whose component fields are the metric $g_{\mu\nu}^{(11)}$, the 3-form tensor $C_{\mu\nu\rho}$, and the gravitino $\psi_\mu$. They sit in irreducible representations of the $SO(9)$ little group, i.e., $\mathbf{44}$, $\mathbf{84}$, and $\mathbf{128}$. We define the index $f_{\rm bulk}$ to be the trace 
 \begin{align}
 	\label{eq:f0}
 	\text{Tr}\Big[(-1)^F \, e^{-\beta \{ Q , Q^{\dagger} \}} \, q^k \, t^{2 (J_{1R} + J_{2R})} u^{2 J_{1L}} v^{2J_{2L}} \Big]
 \end{align}
 over the Kaluza-Klein fields of 11d supergravity on $\mathbb{R}^{1,9} \times S^1$. $k$ is the integer-valued momentum number along the M-circle, such that $|P_0| = |P_{11}| = k/R_M > 0$. Taking into account all Kaluza-Klein modes along the M-circle, the index $f_{\rm bulk}$ takes the form of $f_{\rm bulk} = f_0 \cdot \frac{q}{1-q}$. To compute $f_0$, we combine the index of  particles' translational zero modes on the $\mathbb{R}^8$ (that $SO(8)$ acts on)
\begin{align}
	\label{eq:bosonic-zero-mode}
	\frac{1}{((tu)^{1/2}-(tu)^{-1/2})^{2}((tu^{-1})^{1/2}-(tu^{-1})^{-1/2})^{2}((tv)^{1/2}-(tv)^{-1/2})^{2}((tv^{-1})^{1/2}-(tv^{-1})^{-1/2})^{2}},
\end{align}
and the sum of $SO(8)$ characters $\mathcal{X}_0$ for the following irreducible representations:
\begin{align}	
	\label{eq:11d-sugra-so8-rep}
	\mathbf{44} \oplus \mathbf{84} \oplus \mathbf{128} \rightarrow & \quad (\mathbf{1} \oplus \mathbf{8}_v \oplus \mathbf{35}_v) \oplus (\mathbf{28} \oplus \mathbf{56}_v) \oplus (\mathbf{8}_s \oplus \mathbf{8}_c \oplus \mathbf{56}_s \oplus \mathbf{56}_c)\\
	=& \quad (\mathbf{1} \oplus \mathbf{28} \oplus \mathbf{35}_v) \oplus (\mathbf{8}_v \oplus \mathbf{56}_v)  \oplus (\mathbf{8}_s \oplus \mathbf{56}_s) \oplus (\mathbf{8}_c \oplus \mathbf{56}_c) \nn \\ 
	=& \quad (\mathbf{8}_v  \otimes \mathbf{8}_v ) \oplus (\mathbf{8}_s \otimes \mathbf{8}_c) \oplus (\mathbf{8}_v \otimes \mathbf{8}_c) \oplus (\mathbf{8}_s \otimes \mathbf{8}_v).\nn
\end{align}
Taking $(-1)^F$ into account, $\mathcal{X}_0$ is given by 
\begin{align}
	\mathcal{X}_0 = (\chi(\mathbf{8}_v) - \chi(\mathbf{8}_s)) \cdot  (\chi(\mathbf{8}_v) - \chi(\mathbf{8}_c)) 
\end{align}
where $\chi(\mathbf{R})$ denotes a character for an irreducible representation $\mathbf{R}$. Using the fugacities $t$, $u$, $v$ introduced in \eqref{eq:def-witten-index}, the $SO(8)$ characters $\chi(\mathbf{8}_v)$, $\chi(\mathbf{8}_s)$, and $\chi(\mathbf{8}_c)$ are expressed as follows:
\begin{align}
    \mathbf{8}_v = (\mathbf{2},\mathbf{2},\mathbf{1},\mathbf{1}) \oplus (\mathbf{1},\mathbf{1},\mathbf{2},\mathbf{2}) \longrightarrow &\quad \chi(\mathbf{8}_v) = (t+t^{-1}) \left((u+u^{-1}) + (v+v^{-1})\right) \nonumber\\
    \label{eq:so8-character}
    \mathbf{8}_s = (\mathbf{2},\mathbf{1},\mathbf{2},\mathbf{1}) \oplus (\mathbf{1},\mathbf{2},\mathbf{1},\mathbf{2}) \longrightarrow &\quad \chi(\mathbf{8}_s) = (t+t^{-1})^2 + (u+u^{-1})(v+v^{-1}) \\
    \mathbf{8}_c = (\mathbf{2},\mathbf{1},\mathbf{1},\mathbf{2}) \oplus (\mathbf{1},\mathbf{2},\mathbf{2},\mathbf{1}) \longrightarrow &\quad \chi(\mathbf{8}_c) = (t+t^{-1}) \left((u+u^{-1}) + (v+v^{-1})\right). \nonumber
\end{align}
$\chi(\mathbf{8}_v) =  \chi(\mathbf{8}_c)$ gives $\mathcal{X}_0 = f_\text{0} = 0$. It explains why $Z_\text{ADHM}^\text{U(0)} = 1$, implying $Z_\text{extra}^\text{U(N)}=1$ at the same time.

\vspace{-0.5\baselineskip}
\paragraph{\underline{D0-O4${}^-$}} The O4${}^-$ background uplifts to M-theory on $\mathbb{R}^{1,4} \times \mathbb{R}^{5}/\mathbb{Z}_2 \times S^1$ \cite{Hori1999a}. The $\mathbb{Z}_2$ action inverts the coordinates $(x^5,\cdots, x^9) \, \rightarrow \, (-x^5, \cdots, -x^9)$ and flips the sign of the 3-form tensor $C_3 \rightarrow -C_3$. One can divide the $\mathbb{Z}_2$ action into two operations: (1) the rotation of $x^{5,6}$ and $x^{7,8}$ planes by $+\pi$ and $-\pi$, (2)  the $x^9 \rightarrow -x^9$ parity along with the sign flip of the 3-form tensor $C_3 \rightarrow -C_3$. Regarding the index computation, the first operation $\mathcal{P}_4$ causes the shift of chemical potential $m \rightarrow  m + i \pi$ conjugate to $2J_\text{2L} \equiv J_3 - J_4$, where $J_3$ and $J_4$ are the rotation generators for $x^{5,6}$ and $x^{7,8}$ planes. The second operation $\mathcal{P}_1$ provides the grading of all $SO(8)$ representations in \eqref{eq:11d-sugra-so8-rep}. The grading rules are stated as follows. When $\mathcal{P}_1$ acts,
\vspace{-0.5\baselineskip}
\begin{itemize}
\setlength\itemsep{-0.1em}
	\item Each vector index referring the $x^9$ direction yields the negative sign $(-1)$.
	\item Each spinor index undergoes the multiplication by $\Gamma^9$ where $\Gamma^\mu$ denotes the 11d gamma matrix.
	\item The 3-form tensor $C_3$ goes through the extra sign flip $C_3 \rightarrow -C_3$. 
\end{itemize}
\vspace{-0.5\baselineskip}
We require $\mathbf{8}_s$ to be even and $\mathbf{8}_c$ to be odd under the multiplication by $\Gamma^9$, such that the supercharges $Q$ and $Q^\dagger$ in $\mathbf{8}_s$ which used to define the index \eqref{eq:def-witten-index} are invariant. $\mathbf{56}_c$ and $\mathbf{56}_s$ are parts of $\mathbf{8}_v \otimes \mathbf{8}_s$ and $\mathbf{8}_v \otimes \mathbf{8}_c$, inheriting the parities of $\mathbf{8}_s$ and $\mathbf{8}_c$. 

One can decompose the 11d metric $g_{\mu\nu}^{(11)}$($\mathbf{44}$) into $g_{ij}^{(11)}$($\mathbf{35}_v$), $g_{9i}^{(11)}$($\mathbf{8}_v$), $g_{99}^{(11)}$($\mathbf{1}$), the 3-form tensor $C_{\mu\nu\rho}$ ($\mathbf{84}$) into $C_{ijk}$ ($\mathbf{56}_v$) and $C_{9ij}$ ($\mathbf{28}$), the spin-$\frac{3}{2}$ gravitino $\psi_\mu$ into $\psi_{i}$ ($\mathbf{56}_s \oplus \mathbf{56}_c$) and $\psi_{9}$ ($\mathbf{8}_s \oplus \mathbf{8}_c$). \linebreak  All parity-odd fields are listed as follows: $g_{9i}^{(11)}$($\mathbf{8}_v$),  $C_{ijk}$ ($\mathbf{56}_v$), $\frac{1+\Gamma^9}{2} \cdot \psi_{9}$ ($\mathbf{8}_s$), $\frac{1-\Gamma^9}{2} \cdot \psi_{i}$ ($\mathbf{56}_s$). We define the index $f_\text{bulk}$ as the trace over the Kaluza-Klein fields of 11d supergravity on $\mathbb{R}^{1,4} \times \mathbb{R}^{5}/\mathbb{Z}_2 \times S^1$ with insertion of the projection operator $\frac{1 + \mathcal{P}_4\cdot\mathcal{P}_1}{2}$. Again taking into account all Kaluza-Klein modes along the M-circle, the index $f_\text{bulk}$ takes the form of $\frac{1}{2} (f_0\cdot \frac{q}{1-q} + f_0' \cdot \frac{q}{1-q})$ where $f_0 \cdot \frac{q}{1-q}$ is the index defined in \eqref{eq:f0}. The trace with the parity operator $\mathcal{P}_4\cdot\mathcal{P}_1$
 \begin{align}
 	\label{eq:f0p}
 	\text{Tr}\Big[(-1)^F\, \mathcal{P}_4 \cdot \mathcal{P}_1 \, e^{-\beta \{ Q , Q^{\dagger} \}} \, q^k \, t^{2 (J_{1R} + J_{2R})} u^{2 J_{1L}} v^{2J_{2L}} \Big]
 \end{align}
  gives  $f_0' \cdot \frac{q}{1-q}$.
 $f_0'$ is the product of \eqref{eq:bosonic-zero-mode} and the sum of $SO(8)$ characters
\begin{align}
	\mathcal{X}_1 = &+\{ \chi(\mathbf{1}) - \chi(\mathbf{8}_v)  + \chi(\mathbf{35}_v) + \chi(\mathbf{28}) - \chi(\mathbf{56}_v)\} - \{ \chi(\mathbf{8}_c) + \chi(\mathbf{56}_c) - \chi(\mathbf{56}_s) -\chi(\mathbf{8}_s) \}\nn \\
	&\nn = \chi(\mathbf{8}_v)^2 - \chi(\mathbf{8}_s) \chi(\mathbf{8}_c) - \chi(\mathbf{8}_v) \chi(\mathbf{8}_s) + \chi(\mathbf{8}_c) \chi(\mathbf{8}_v) = (\chi(\mathbf{8}_v) + \chi(\mathbf{8}_s)) \cdot (\chi(\mathbf{8}_v) - \chi(\mathbf{8}_c))
\end{align}
where $\mathcal{P}_4$ replaces $v \equiv e^{-m}$ by $-v = e^{-m - i\pi}$ in \eqref{eq:bosonic-zero-mode} and $\mathcal{X}_1$. After all, the index $f_\text{bulk}$ becomes
\begin{align}
	 f_\text{bulk} = \frac{t^2 (v +v^{-1} -u - u^{-1} )(t + t^{-1})}{(1-tu)(1-tu^{-1})(1 + tv)(1 + tv^{-1})} \frac{q}{1-q}.
\end{align}
It shows  $Z_\text{ADHM}^\text{SO(0)}  = \text{PE}\;[\frac{1}{2}\,f_\text{bulk}]$ in \eqref{eq:decoupled-so2n} counts the 11d supergravity fields on $\mathbb{R}^{1,4} \times \mathbb{R}^5 / \mathbb{Z}_2 \times S^1$, up to the $\frac{1}{2}$ factor, which do not belong to the 5d SYM Hilbert space. We conclude that $Z_\text{extra}^\text{SO(2N)} = Z_\text{ADHM}^\text{SO(0)}$.

\vspace{-0.5\baselineskip}
\paragraph{\underline{D0-O4${}^0$}}  The O4${}^0$ background uplifts to M-theory on $\mathbb{R}^{1,4} \times (\mathbb{R}^{5} \times S^1)/\mathbb{Z}_2$, whose $\mathbb{Z}_2$ action is the combination of $\mathcal{P}_4$, $\mathcal{P}_1$, and the M-circle shift $x^{11} \rightarrow x^{11} + \pi R_M$ by the half-period \cite{Hori1999a}. In addition to the $\mathcal{P}_4$ and $\mathcal{P}_1$ actions described above, the M-circle shift induces the phase factor $e^{i\pi} = -1$ to a single D0-brane wavefunction $\exp{(i x^{11}/R_M)}$. It causes the sign change of the momentum fugacity $q \rightarrow -q$. The index $f_\text{bulk}$ over the Kaluza-Klein fields of 11d supergravity on $\mathbb{R}^{1,4} \times (\mathbb{R}^{5} \times S^1)/\mathbb{Z}_2$ becomes
\begin{align}
	 f_\text{bulk} = \frac{t^2 (v +v^{-1} -u - u^{-1} )(t + t^{-1})}{(1-tu)(1-tu^{-1})(1 + tv)(1 + tv^{-1})} \frac{-q}{1-(-q)},
\end{align}
whose multi-particle spectrum agrees with $Z_\text{ADHM}^\text{SO(1)}$ in \eqref{eq:decoupled-so2nodd} up to the $\frac{1}{2}$ factor. This again explains that $Z_\text{ADHM}^\text{SO(1)}$ comes from the 11d supergravity, implying that $Z_\text{extra}^\text{SO(2N+1)} = Z_\text{ADHM}^\text{SO(1)}$.

\vspace{-0.5\baselineskip}
\paragraph{\underline{D0-O4${}^+$}} The O4${}^+$ background uplifts to M-theory on $\mathbb{R}^{1,4} \times \mathbb{R}^{5}/\mathbb{Z}_2 \times S^1$ with the full M5-brane frozen at the $\mathbb{Z}_2$ fixed plane \cite{Hori1999a}. $q$ is the fugacity for half-integral momenta along the M-circle. The first term of $Z_\text{ADHM}^\text{Sp(0)${}_{\theta = 0}$}$ in \eqref{eq:decoupled-spn-notheta} comes from the 11d bulk gravity, using the same argument used for the O4${}^-$ background. Here we take into account that $q^2$ is the fugacity for integral momenta in the bulk. We identify this as the extra states' index $Z_\text{extra}^\text{Sp(N)${}_{\theta = 0}$}$, such that
\begin{align}
	Z_\text{extra}^\text{Sp(N)${}_{\theta = 0}$} = \text{PE}\left[\frac{t^2 (v+v^{-1}-u-u^{-1})(t+t^{-1})}{2(1-tu^\pm)(1+tv^\pm)}\frac{q^2}{1-q^2} \right].
\end{align}
The second term in \eqref{eq:decoupled-spn-notheta} comes from the full M5-brane frozen at the $\mathbb{Z}_2$ fixed plane, which hosts the $SO(2)$-type $(2,0)$ theory on $T^2$ with the $\mathbb{Z}_2$ outer automorphism twist \cite{Tachikawa2011b}.  Beginning from the Abelian $(2,0)$ theory on $S^1$, whose index is given by
\begin{align}
	\text{PE}\left[\frac{t (v+v^{-1}-u-u^{-1})}{(1-t u)(1-t u^{-1})}\frac{q^2}{1-q^2} \right] \quad \text{with $q =e^{\pi i\tau}$},
\end{align}
the result in \eqref{eq:decoupled-spn-notheta} shows that all states with integral momenta are eliminated by the $\mathbb{Z}_2$ projection, while the twisted states having half-integral momenta are newly introduced. The perturbative degrees for 5d Abelian SYM are projected out, yielding the pure orientifold background. Our result is consistent with \cite{Keurentjes2002} which shows a half D0-brane can bind to an O4${}^+$ plane while a full D0-brane do not.

Applying the S-duality of M-theory on $T^2$, which swaps the temporal circle and the M-circle, the O4${}^+$ background is mapped to the full D4-brane frozen at the O4${}^-$ plane \cite{Hori1999a}. In the S-dualized background, the $O(2)_-$ Wilson line $\sigma_3 = \text{diag}\,(+1, -1)$ along the temporal circle prevents the D4-brane from moving away from the orientifold \cite{Gimon1998}. We shall verify the S-duality relation between two orientifold backgrounds using the multi-particle indices of the relevant ADHM quantum mechanices. First, we compute the 5d $O(2)$ instanton partition function in the Wilson line background $\sigma_3 = \text{diag}\,(+1, -1) \in O(2)_-$. It can be done by replacing the continuous holonomies $\pm \alpha_1$ of $G = SO(2)$ by the discrete holonomies $0$ and $i\pi$ of $O(2)_-$ in the 1-loop determinants $I_\text{1-loop}$. Such replacement means the full D4-brane cannot move from the O4${}^-$ plane. Following the computation in Section~\ref{subsec:witten-index}, the index $Z^{O(2)_-}$ is expressed using the plethystic exponential as follows:
\begin{align}
\label{eq:inst-o2m}
	Z^{O(2)_-} = \text{PE}\, \bigg[&-\frac{t (v+v^{-1}-u-u^{-1})}{(1-tu)(1-tu^{-1})}\frac{q}{1-q^2} -\frac{t^2 (v+v^{-1}-u-u^{-1})(v+v^{-1}-t-t^{-1})}{(1-tu)(1-tu^{-1})(1+tu)(1+tu^{-1})}\frac{q^2}{1-q^2}\nn \\
	& -\frac{t^2 (u + u^{-1} -v -v^{-1})(t + t^{-1})}{2(1-tu)(1-tu^{-1})(1 + tv)(1 + tv^{-1})}\frac{q}{1-q}\bigg]
\end{align}	
whose $q$-dependence was checked up to $q^3$-order. The last term comes from the 11d bulk gravity. The remaining two terms are expected to be S-dual to the second term of $Z_\text{ADHM}^\text{Sp(0)${}_{\theta = 0}$}$.

We observe that the second term of $Z_\text{ADHM}^\text{Sp(0)${}_{\theta = 0}$}$ can be written as 
\begin{align}
	\label{eq:o4+}
	\text{PE}\left[\frac{t (v+v^{-1}-u-u^{-1})}{(1-t u)(1-t u^{-1})}\frac{q}{1-q^2}\right]  = \frac{\mathcal{Z}(\frac{\tau}{2}, \epsilon_+, \epsilon_-, m)}{\mathcal{Z}(\tau, \epsilon_+, \epsilon_-, m)} \quad \text{with $q =e^{\pi i\tau}$}
\end{align}
where 
\begin{align}
	\mathcal{Z}(\tau,\epsilon_+, \epsilon_-, m) \equiv \text{PE}\left[\frac{t (v+v^{-1}-u-u^{-1})}{(1-t u)(1-t u^{-1})}\bigg(\frac{q^2}{1-q^2} + \frac{1}{2}\bigg)\right] \quad \text{with $q =e^{\pi i\tau}$}.
\end{align}
 The S-duality transformation of this function has been studied in \cite{Kim2013a}, since 
$\mathcal{Z}$ is the partition function of a single M5-brane on Omega-deformed $\mathbb{R}^4\times T^2$. Let us apply the S-duality transformation that exchanges the two sides of the $T^2$, i.e., $\tau \rightarrow -\frac{1}{\tau}$, and transforms the chemical potentials such that $\epsilon_+ \rightarrow \tfrac{\epsilon_+}{\tau}$, $\epsilon_- \rightarrow \tfrac{\epsilon_-}{\tau}$, and $m \rightarrow \tfrac{m}{\tau}$. The S-dual partition function $\mathcal{Z}(-\frac{1}{\tau}, \tfrac{\epsilon_+}{\tau}, \tfrac{\epsilon_-}{\tau}, \tfrac{m}{\tau})$ is related to the original one in the following simple manner \cite{Kim2013a},
\begin{align}
	\label{eq:s-dual-inv}
	\mathcal{Z}(\tau, \epsilon_+, \epsilon_-, m)  = \exp{\left(\tfrac{2\pi i\, (m^2-\epsilon_-^2)(m^2-\epsilon_+^2)}{24 \tau\,  (\epsilon_+ + \epsilon_-) (\epsilon_+ - \epsilon_-)}\right)} \  \mathcal{Z}(-\tfrac{1}{\tau}, \tfrac{\epsilon_+}{\tau}, \tfrac{\epsilon_-}{\tau}, \tfrac{m}{\tau}),
\end{align}
in a particular parameter regime. See \cite{Kim2013a} for the details. $\mathcal{Z}(\tau, \epsilon_+, \epsilon_-, m)$ is essentially invariant under S-duality, up to a simple overall factor \cite{Kim2013a}. Accordingly, \eqref{eq:o4+} becomes
\begin{align}
	\frac{\mathcal{Z}(2\tau, 2\epsilon_+, 2\epsilon_-, 2m)}{\mathcal{Z}(\tau, \epsilon_+, \epsilon_-, m)} =&\ \text{PE}\left[\frac{t^2 (v^2+v^{-2}-u^2-u^{-2})}{(1-t^2 u^{\pm 2})}\bigg(\frac{q^2}{1-q^2}+\frac{1}{2}\bigg) -\frac{t (v+v^{-1}-u-u^{-1})}{(1-t u^\pm)}\bigg(\frac{q}{1-q}+\frac{1}{2}\bigg)\right]\nn \\
	= &\ \text{PE}\, \bigg[-\frac{t (v+v^{-1}-u-u^{-1})}{(1-tu)(1-tu^{-1})}\frac{q}{1-q^2} -\frac{t^2 (v+v^{-1}-u-u^{-1})(v+v^{-1}-t-t^{-1})}{(1-tu)(1-tu^{-1})(1+tu)(1+tu^{-1})}\frac{q^2}{1-q^2}\bigg]&&\nn \\ 
	\label{eq:s-dualized-o4+}
	\times &\  \text{PE}\, \bigg[ -\frac{t^2 (v+v^{-1}-u-u^{-1})(v+v^{-1}-t-t^{-1})}{2(1-tu)(1-tu^{-1})(1+tu)(1+tu^{-1})}\bigg]
\end{align}
under the S-duality transformation.
The second line equals the first line of $Z^{O(2)_-}$ in \eqref{eq:inst-o2m}. Apart from the $\tau$ independent factor on the third line of \eqref{eq:s-dualized-o4+}, one finds that $Z_\text{ADHM}^\text{Sp(0)${}_{\theta = 0}$}$ is dual to $Z^{O(2)_-}$. This analysis confirms the S-duality relation studied in \cite{Gimon1998} between the O4${}^+$ plane and the O4${}^-$ plane with the frozen D4-brane.

\vspace{-0.5\baselineskip}
\paragraph{\underline{D0-$\widetilde{\text{O4}}^+$}} The $\widetilde{\text{O4}}^+$ background uplifts to M-theory on $\mathbb{R}^{1,4} \times (\mathbb{R}^5 \times S^1)/\mathbb{Z}_2$ with a stuck M5-brane at the origin of $\mathbb{R}^5$ \cite{Hori1999a}. The $\mathbb{Z}_2$ action consists of the $\mathbb{R}^5$ parity and the M-circle shift $x^{11} \rightarrow x^{11} + \pi R_M$ by the half-period. The M-circle shift induces the phase factor $e^{i\pi/2} = i$ to a half D0-brane wavefunction $\exp{(ix^{11}/2R_M)}$, causing the half-integral momentum fugacity $q$ to undergo the phase rotation $q \rightarrow i q$. Repeating the same argument used for the O4${}^0$ background, we understand the first term of $Z_\text{ADHM}^\text{Sp(0)${}_{\theta = \pi}$}$ as coming from the 11d bulk gravity. It implies that $Z_\text{extra}^\text{Sp(N)${}_{\theta = \pi}$}$ should be identified as
\begin{align}
	Z_\text{extra}^\text{Sp(N)${}_{\theta = \pi}$} = \text{PE}\left[\frac{t^2 (v+v^{-1}-u-u^{-1})(t+t^{-1})}{2(1-tu^\pm)(1+tv^\pm)}\frac{-q^2}{1-(-q^2)}\right].
\end{align}
The second term of $Z_\text{ADHM}^\text{Sp(0)${}_{\theta = \pi}$}$ in \eqref{eq:decoupled-spn-theta} is similar to that of $Z_\text{ADHM}^\text{Sp(0)${}_{\theta = 0}$}$ in \eqref{eq:decoupled-spn-notheta}. Both are obtained from single M5-branes stuck at the $\mathbb{Z}_2$ fixed point of $\mathbb{R}^5$. The difference is that the M5-brane wrapping the freely-acting $\mathbb{Z}_2$ orbifold sees the halved M-circle radius. It effectively doubles the complex structure $\tau$ of the $T^2$, explaining the difference between \eqref{eq:o4+} and 
\begin{align}
	\label{eq:o4+theta}
	\text{PE}\left[\frac{t (v+v^{-1}-u-u^{-1})}{(1-t u)(1-t u^{-1})}\frac{q^2}{1-q^4 } \right]   \quad \text{with $q =e^{\pi i\tau}$}.
\end{align}
Similar to the O4${}^+$ case, the $\mathbb{Z}_2$ projection kills all integral-momentum states in the Abelian $(2,0)$ theory on $S^1$. What contribute to \eqref{eq:o4+theta} are the twisted states, having half-integral momenta along the M-theory circle. The $\mathbb{Z}_2$ projection eliminates the perturbative degrees of 5d Abelian SYM, leaving the pure orientifold background. Its S-dual configuration in M-theory has been suggested in \cite{Hori1999a,Gimon1998}. It would be interesting to see the S-duality of the two backgrounds explicitly.

We remark that the multi-particle indices \eqref{eq:decoupled-so2n}-\eqref{eq:decoupled-spn-theta}  of the ADHM quantum mechanics for various D0-O4 systems capture the non-perturbative (curvature)${}^2$ terms of orientifold backgrounds. Recall that the genus expansion of the topological string amplitude $F$ takes the form of \cite{Gopakumar1998a,Gopakumar1998,Iqbal2009}
\begin{align}
	F \equiv \log{Z_{\rm ADHM}} = \sum_{n, g\geq 0}^\infty (\epsilon_1 + \epsilon_2)^{2n} (\epsilon_1\epsilon_2)^{g-1} \mathcal{F}^{(n,g)}.
\end{align}
Turning off $\epsilon_+$, the remaining Omega-deformation parameter $\epsilon_-$ denotes the self-dual part of the graviphoton field strength on $\mathbb{R}^4$. The amplitude gives the (curvature)${}^2$ correction to M-theory on $\text{CY3} \times S^1$ \cite{Gopakumar1998}.  Considering the case with $m=\epsilon_+=0$, the genus-$1$ terms at $n=0$ are given from \eqref{eq:decoupled-so2n}-\eqref{eq:decoupled-spn-theta} as follows.
\begin{align}
	&Z_\text{ADHM}^\text{SO(0)} \rightarrow  -\frac{1}{4} \log\left(\eta(\tau)\right),&  &Z_\text{ADHM}^\text{SO(1)} \rightarrow  -\frac{1}{4}\log\left(\frac{\eta(2\tau)^{2}}{\eta(\tau)}\right)\\ 
	&Z_\text{ADHM}^\text{Sp(0)${}_{\theta = 0}$} \rightarrow  -\frac{1}{4} \log\left(\eta(\tau)\right) - \log\left(\frac{\eta(\tau/2)}{\eta(\tau)}\right) &&
	Z_\text{ADHM}^\text{Sp(0)${}_{\theta = \pi}$} \rightarrow -\frac{1}{4}\log\left(\frac{\eta(2\tau)^{2}}{\eta(\tau)}\right) - \log\left(\frac{\eta(\tau)}{\eta(2\tau)}\right) \nn.
\end{align}
After performing the S-duality transformation $\tau \rightarrow -\frac{1}{\tau}$, they agree with the (curvature)${}^2$ terms in the O4-plane actions, computed from graviton scattering amplitudes with non-perturbative effects \cite{Henry-Labordere2002}, up to an overall factor $-\frac{1}{4\pi}$ that we have not precisely traced.

\section{6d SCFTs from 5d SYM instantons}
\label{sec:6dscft-5dsym}
In this section, we study the BPS spectra of 6d $(2,0)$ theories using the instanton partition functions. String theory predicts that 5d maximal SYM, supplemented by the non-perturbative instanton effect, has a UV fixed point which corresponds to 6d $(2,0)$ SCFT \cite{Douglas2011,Lambert2011}. The instanton partition function $Z_\text{inst}$ would capture the BPS spectrum of 6d $(2,0)$ SCFT wrapped on $\mathbb{R}^4 \times T^2$, where instantons play the role of Kaluza-Klein momenta along the M-circle. For example, the interpretation of the $SU(N)$ instanton partition function as the tensor branch index of 6d $SU(N)$-type $(2,0)$ SCFT was justified in \cite{Kim2011,Haghighat2015a}. We want to extend some aspects of the above analysis to other gauge groups: $SO(2N)$, $SO(2N+1)$, and $Sp(N)$. 

In Section~\ref{subsec:consistency}, we make some consistency checks of the instanton partition functions for lower rank gauge groups. This is because the $SU(N)$, $SO(2N+1)$, $Sp(N)$, $SO(2N)$ Lie algebras sometimes coincide at low ranks. We confirm that apparently different ADHM approaches yield the same prepotentials. In Section~\ref{subsec:s-dual}, we obtain the closed form expressions for the instanton partition functions in special limits, and study the S-duality relations of them. In Section~\ref{subsec:superconformal-w}, we study the 6d superconformal indices using the results of Section~\ref{subsec:s-dual}.

\subsection{Consistency checks}
\label{subsec:consistency}

Let us compare the instanton partition functions with coincident gauge groups, which belong to different ABCD-types of Lie algebras. For instance, the classical actions of the maximal super Yang-Mills theories with $SU(2)$ and $SO(3)$ gauge groups are identical, both in 4d and 5d. In 4d, this means that the two quantum theories are completely the same. So the physical QFT observables that one can extract out of the instanton partition functions should be the same, although the two ADHM descriptions look different. However, in 5d, having identical classical actions do not necessarily imply that the two instanton calculus are the same, because different ADHM quantum mechanics for 5d solitons may provide different UV completions of non-renormalizable 5d Yang-Mills theories.

The requirement that $SU(2)$ and $SO(3)$ partition functions should yield identical 4d observables imposes non-trivial constraints on the two partition functions, which we shall check in this section. One important 4d observable is the prepotential, whose second derivative with the Coulomb VEVs $\alpha$ gives the Coulomb branch effective action. Thus, the Coulomb VEV dependent parts of the $SU(2)$ and $SO(3)$ partition functions should be the same. We check in the 5d partition functions that the Coulomb VEV dependent parts are indeed identical to each other, even before taking the 4d limits, which confirms that the two partition functions satisfy the 4d constraint. However, Coulomb VEV independent parts do not have to agree with each other, and we find that they indeed disagree for $SU(2)$ and $SO(3)$. We make similar consistency checks for gauge theories with low ranks, whenever we have more than one ADHM descriptions: $SO(2)$ vs $U(1)$, $SO(4)$ vs $SU(2)^2$, $SO(6)$ vs $SU(4)$, $SU(2)$ vs $SO(3)$ vs $Sp(1)$, $SO(5)$ vs $Sp(2)$.

However, in 5d, even Coulomb VEV independent terms are physically meaningful, as the partition function itself is a Witten index. They are simply counting electrically neutral BPS states. Different ADHM descriptions may provide different UV completions of the two 5d gauge theories which are identical at the classical level. Such phenomena happen when the 5d gauge groups belong to the B or C classes, which uplift to different 6d theories on a circle with outer automorphism twists. It also affects the 5-sphere partition functions that we study in Section~\ref{subsec:superconformal-w}, which we interpret as 6d superconformal indices with outer automorphism twists. $SO(2)$ vs $U(1)$, $SO(4)$ vs $SU(2)^2$, $SO(6)$ vs $SU(4)$ theories, which formally belong to the AD classes, are expected to have identical UV uplifts. In these cases, we expect that both the charged and neutral sectors are the same.

We start by commenting on our parameterizations of the $SU(N)$ chemical potentials. The $SU(N)$ instanton partition functions can be obtained from the $U(N)$ instanton partition functions by imposing the traceless conditions on the chemical potentials. Among the $U(N)$ holonomies $\alpha_{1}$, $\cdots$, $\alpha_{N}$, we turn off the overall $U(1)$ holonomy $\sum_{i=1}^N \alpha_i = 0$ and replace all $\alpha_{i}$ by the $SU(N)$ holonomies $\alpha_i'$ defined by $\alpha_i' \equiv \alpha_{i+1} - \alpha_i$. The final expressions depend on $\alpha_1'$, $\cdots$, $\alpha_{N-1}'$.

\vspace{-\baselineskip}
\paragraph{\underline{SO(2) vs U(1)}} The instanton partition functions for Abelian theories are, checked up to $q^3$ order,
\begin{align}
	Z^\text{SO(2)}_\text{inst} = Z^\text{U(1)}_\text{inst} = \text{PE} \left[\frac{t(v+v^{-1}-u-u^{-1})}{(1-tu)(1-tu^{-1})}\frac{q}{1-q}\right].
\end{align}
They are independent of the gauge holonomies, since the Abelian theories do not involve electrically charged particles. They agree with the 6d index on $\mathbb{R}^4 \times T^2$ for a free $(2,0)$ tensor multiplet \cite{Kim2011}.

\vspace{-\baselineskip}
\paragraph{\underline{SO(4) vs SU(2)${}^2$}}  The $SO(4)$ and $SU(2)^2$ instanton partition functions satisfy the following relation
\begin{align}
	\label{eq:so4-su2-su2}
	Z^\text{SO(4)}_\text{inst} (w_1, w_2, q) = Z^\text{SU(2)}_\text{inst} (w_1' = w_1 w_2, q) \cdot Z^\text{SU(2)}_\text{inst} (w_1' = w_1^{-1} w_2, q).
\end{align}
which was checked by taking the series expansion in $w_{1,2}$,  up to $w_1^3\, w_2^3 \, q^2$ order.

The $SO(4)$ instanton partition function depends on the $SO(4)$ holonomies $\alpha_1$ and $\alpha_2$, which are distances of two D4-branes from the orientifold. There are two kinds of charged W-bosons in the $SO(4)$ gauge theory, induced from open strings whose lengths are $|\alpha_1 \pm \alpha_2|$. The relation \eqref{eq:so4-su2-su2} shows that each type of W-bosons is that of the $SU(2)$ gauge theory.

\vspace{-\baselineskip}
\paragraph{\underline{SO(6) vs SU(4)}}  The $SO(6)$ and $SU(4)$ instanton partition functions satisfy the following relation
\begin{align}
	\label{eq:so6-su4}
	Z^\text{SO(6)}_\text{inst} (w_1, w_2, w_3, q) = Z^\text{SU(4)}_\text{inst} (w_1' = w_1 w_2, w_2' = w_1^{-1} w_2, w_3' = w_2^{-1}w_3, q) 
\end{align}
which was checked by taking the series expansion in $w_{1,2,3}$, up to $w_1^1\, w_2^1 \, w_3^1\, q^2$ order.

The $SO(6)$ gauge holonomies $\alpha_{1,2,3}$, measuring the distances of three D4-branes from the orientifold, appear in the $SO(6)$ instanton partition function. There are three types of charged W-bosons in the $SO(6)$ gauge theory, induced from open strings whose lengths are $|\alpha_1 \pm \alpha_2|$ and $|\alpha_2 - \alpha_3|$. They correspond to three kinds of $SU(4)$ W-bosons, according to the relation \eqref{eq:so6-su4}.

\vspace{-\baselineskip}
\paragraph{\underline{SU(2) vs SO(3) vs Sp(1)$|_{\theta=0}$}} The instanton partition functions for these theories are written as
\begin{align}
	\label{eq:su2-inst}
	Z_\text{inst}^{SU(2)} &= \text{PE}\left[f^{\rm SU(2)} (w,q) + \frac{t(v+v^{-1}-u-u^{-1})}{(1-tu)(1-tu^{-1})}\frac{q}{1-q}\right] && \text{with } w = e^{-\alpha'_1}\\
	Z_\text{inst}^{SO(3)} &= \text{PE}\left[f^{\rm SO(3)} (w, q) + \frac{t(v+v^{-1}-u-u^{-1})}{(1-tu)(1-tu^{-1})}\frac{q}{1-q} \right] && \text{with } w = e^{-\alpha_1}\\
	Z_\text{inst}^{Sp(1)} &= \text{PE}\left[f^{\rm Sp(1)} (w, q) + \frac{t(v+v^{-1}-u-u^{-1})}{(1-tu)(1-tu^{-1})}\frac{q}{1-q}\right] && \text{with } w = e^{-2\alpha_1}
\end{align}
where $\alpha'_1 = \alpha_1 - \alpha_2$.
We divide the instanton partition functions into two parts: The first term $f$ is dependent on the Coulomb VEVs. The second term is independent of the Coulomb VEVs. 

We find that the 4d QFT constraint is satisfied, $f^{\rm SU(2)}(w, q) = f^{\rm SO(3)}(w, \sqrt{q}) = f^{\rm Sp(1)}(w,q)$, checked up to $q^2 w^3$ order. However, the Coulomb VEV independent term of the $SO(3)$ theory disagrees with those of the $SU(2)$ and $Sp(1)$ theories, using the same identification of $q$'s. It implies that these gauge theories do not have the same UV completion. The $Sp(1)$ gauge theory with $\theta=\pi$ should satisfy the same constraint, which can be straightforwardly checked but we have not checked.

The four 5d theories with $SU(2)$, $SO(3)$, $Sp(1)_{\theta=0}$, $Sp(1)_{\theta=\pi}$ uplift to 6d $(2,0)$ theories of $A_1$-type, $A_1$-type with $\mathbb{Z}_2$ outer automorphism twist, $D_2$-type with $\mathbb{Z}_2$ outer automorphism twist, and $A_2$-type with $\mathbb{Z}_2$ outer automorphism twist, respectively \cite{Tachikawa2011b}. We expect  the differences in the Coulomb VEV independent parts to reflect such distinct UV completions.

\vspace{-\baselineskip}
\paragraph{\underline{SO(5) vs Sp(2)$|_{\theta=0}$}} The instanton partition functions for $SO(5)$ and $Sp(2)$ gauge theories are
\begin{align}
	&Z_\text{inst}^{SO(5)} = \text{PE}\left[f^{\rm SO(5)} (\hat{w}_{1,2},q) + \frac{2t(v+v^{-1}-u-u^{-1})}{(1-tu)(1-tu^{-1})}\frac{q}{1-q} \right] \text{with } \hat{w}_1 = e^{-\alpha_1},\ \hat{w}_2 = e^{-(\alpha_2 - \alpha_1)}\\	
	&Z_\text{inst}^{Sp(2)} = \text{PE}\left[f^{\rm Sp(2)} (\hat{w}_{1,2},q) + \frac{t(v+v^{-1}-u-u^{-1})}{(1-tu)(1-tu^{-1})}\left(\frac{q}{1-q} + \frac{q^2}{1-q^2}\right)\right] \\&
\hspace{10cm}\text{ with } \hat{w}_1 = e^{-2\alpha_1},\ \hat{w}_2 = e^{-(\alpha_2 - \alpha_1)}.\nn
\end{align}
Again we divide the instanton partition functions into the indices from Coulomb VEV dependent sectors and independent sectors. The Coulomb VEV dependent parts satisfy the 4d QFT constraint $f^{\rm SO(5)}(\hat{w}_1, \hat{w}_2, q) = f^{\rm Sp(2)}(\hat{w}_2, \hat{w}_1, q)$ after exchanging $\hat{w}_1 \leftrightarrow \hat{w}_2$. However, the neutral parts do not agree with each other, implying that the two theories uplift to the different 6d quantum field theories. We have not considered the $Sp(2)$ partition function with $\theta=\pi$, but the same analysis can be made.

The three 5d theories with $SO(5)$, $Sp(2)_{\theta=0}$, $Sp(2)_{\theta=\pi}$ uplift to 6d theories of $A_3$-type with $\mathbb{Z}_2$ outer automorphism twist, $D_3$-type with $\mathbb{Z}_2$ outer automorphism twist, and $A_4$-type with $\mathbb{Z}_2$ outer automorphism twist, respectively.

\subsection{Index on $\mathbb{R}^4 \times T^2$ and S-duality}
\label{subsec:s-dual}

Now we study if the instanton partition functions respect S-dualities of 4d $\mathcal{N}=4$ gauge theories, and their 6d uplifts on tori. S-duality identifies certain pairs of 4d $\mathcal{N}=4$ gauge theories, such as \cite{Goddard1977,Montonen1977,Witten1978,Osborn1979}
\begin{align}
	U(N)  &\quad \longleftrightarrow \quad U(N)  &&  \hspace{-0cm}\text{ with  }\ n_G = 1\nn \\
	SO(2N)   &\quad \longleftrightarrow \quad SO(2N) && \hspace{-0cm}\text{ with  }\ n_G = 1 \\
	SO(2N+1)   &\quad \longleftrightarrow \quad  Sp(N) && \hspace{-0cm}\text{ with  }\ n_G = 2, \nn
\end{align}
if their gauge couplings $\tau_4 = \frac{\theta_4}{2\pi} + \frac{4\pi i}{g_4^2}$ and $\tau_4^\vee = \frac{\theta_4^\vee}{2\pi} + \frac{4\pi i}{g_4^{\vee 2}}$ are related as $\tau_4^\vee = -\frac{1}{n_G \tau_4}$. Recall that 4d $\mathcal{N}=4$ SYMs are obtained from $(2,0)$ theories on tori, whose complex structures $\tau\equiv i \frac{\beta/2\pi}{R_M}$ are translated to  gauge couplings $\tau_4$ \cite{Witten1995b}. The precise relations between complex structures $\tau$ and gauge couplings $\tau_4$ are given in \cite{Kapustin2006}. From this viewpoint, S-dualities merely exchange two sides of the tori. 


Our instanton partition functions are six-dimensional observables. However, they depend only on the complex structures $\tau $ of the tori, and not on their volumes. So these partition functions are expected to respect the geometric S-dualities in certain forms. For instance, the S-duality transformation of the partition function for the Abelian $(2,0)$ theory has been studied in \cite{Kim2013a}, as we reviewed in Section~\ref{sec:extra}. Even in such simple theory, the transformation of the 6d partition function is fairly non-trivial. For general non-Abelian theories, we do not know the S-duality transformations of the partition functions, keeping all the chemical potentials. However, the partition functions may simplify after taking special limits of the chemical potentials, so that we can explicitly see the S-duality of the partition functions. 

We are interested in two different limits of the partition functions, which are closely related to each other. One is to take all the fugacities $w_i$ for the $U(1)^N$ electric charges to be zero. In this limit, the partition function acquires contributions only from the neutral states. Another limit we are interested in is $m=\epsilon_+$. This limit also makes the contributions from charged sectors automatically vanish. It is because W-bosons always carry a factor of $2\sinh\frac{m\pm\epsilon_+}{2}$ in their indices, coming from the supersymmetries broken by these BPS states \cite{Kim2011}. Since this factor vanishes at $m=\epsilon_+$, the index acquires contributions only from neutral instantons unbound to W-bosons. The 6d partition functions in both limits can be easily dualized to test the S-dualities suggested in the literatures.

We first consider the limit $w_i\rightarrow 0$. For various gauge groups, we observe that the instanton partition functions in electrically neutral sectors are given as
\begin{align*}
	Z_\text{inst}^{SO(2N)}|_{w_i = 0} &= \text{PE} \left[\frac{N\,t(v+v^{-1}-u-u^{-1})}{(1-tu)(1-tu^{-1})}\frac{q}{1-q}\right]  && \text{checked for $N \leq 6$, up to $q^3$ order}\nn\\
	Z_\text{inst}^\text{SO(2N+1)}|_{w_i = 0} &= \text{PE} \left[\frac{N\,t(v+v^{-1}-u-u^{-1})}{(1-tu)(1-tu^{-1})}\frac{q}{1-q}\right] && \text{checked for $N \leq 6$, up to $q^2$ order}\nn\\
	Z_\text{inst}^\text{Sp(N)$|_{\theta=0}$}|_{w_i = 0} &= \text{PE} \left[\frac{t(v+v^{-1}-u-u^{-1})}{(1-tu)(1-tu^{-1})}\left(\frac{N \, q^2}{1-q^2} + \frac{q}{1-q^2}\right)\right] && \text{checked for $N \leq 3$,  up to $q^5$ order}\nn \\
	Z_\text{inst}^\text{Sp(N)$|_{\theta=\pi}$}|_{w_i = 0} &= \text{PE} \left[\frac{t(v+v^{-1}-u-u^{-1})}{(1-tu)(1-tu^{-1})} \left(\frac{N \, q^2}{1-q^2} + \frac{q^2}{1-q^4}\right)\right] && \text{checked for $N \leq 3$,  up to $q^5$ order}.
\end{align*}
All of these partition functions include the 6d index on $\mathbb{R}^4 \times T^2$ for $N$ free $(2,0)$ tensor multiplets, whose S-duality transformations were studied in \cite{Kim2013a}. Extra terms in the $Sp(N)$ partition functions are already discussed in Section~\ref{sec:extra}.

We are further interested in the special limit $m \rightarrow \epsilon_+$ where the instanton partition functions simplify into the following forms,
\begin{align}
	Z_\text{inst}^{SO(2N)}|_{m = \epsilon_+} &= \text{PE} \left[\frac{N\, q}{1-q}\right] = \eta(\tau)^{-N},\\
	Z_\text{inst}^{SO(2N+1)}|_{m = \epsilon_+} &= \text{PE} \left[\frac{N\, q}{1-q}\right] \; \,= \eta(\tau)^{-N},\\
	Z_\text{inst}^{Sp(N)|_{\theta = 0}}|_{m = \epsilon_+} &= \text{PE} \left[\frac{N\, q^2}{1-q^2} + \frac{q}{1-q^2}\right] = \eta(\tau)^{-N+1} \, \eta(2\tau)^{-1},\\
	Z_\text{inst}^{Sp(N)|_{\theta = \pi}}|_{m = \epsilon_+} &= \text{PE} \left[\frac{N\, q^2}{1-q^2} + \frac{q^2}{1-q^4}\right] = \eta(\tau)^{-N-1} \,\eta(2\tau)^{+1},
\end{align}
up to the overall factors $e^{-N \pi i \tau /12}$ and $e^{-(N \pm 1) \pi i \tau /12}$. These expressions will be useful to study the 6d superconformal indices in Section~\ref{subsec:superconformal-w}. The $SO(2N)$ partition functions are invariant under the S-duality $\tau \rightarrow -\frac{1}{\tau}$  as expected. We expect the other two $Sp(N)$ partition functions to be S-dualized to the partition functions of the 5d gauge theories  on a circle with twisted boundary conditions \cite{Tachikawa2011b}. It will be interesting to explicitly study the dual partition functions with twisted boundary conditions.

\subsection{6d superconformal index and $\mathcal{W}$ algebra}
\label{subsec:superconformal-w}

Here we study the $S^5$ partition functions that correspond to the $(2,0)$ superconformal indices \cite{Kallen2012,Kim2013b,Lockhart2012,Kim2012}. The superconformal indices \cite{Bhattacharya2008} are the SUSY partition functions on $S^5 \times S^1$ which capture the 6d BPS operator spectra, due to the operator-state correspondence of radially quantized CFTs.

It was proposed in \cite{Lockhart2012,Kim2012,Qiu2013} that the $S^5$ partition functions are computed by merging three Nekrasov partition functions on $\mathbb{R}^4 \times S^1$. They depend on the dimensionless coupling $\sigma = \frac{g_5^2}{2\pi r_5}$, made of the $S^5$ radius $r_5$ and the Yang-Mills coupling $g_5$. $\sigma$ can be interpreted as the chemical potential analogous to inverse temperature on $S^5 \times S^1$. The expressions in \cite{Lockhart2012,Kim2012,Qiu2013} take the form of weak coupling expansions ($\sigma \ll 1$), while the 6d spectral data are easier to read off in the strong coupling regime ($\sigma \gg 1$) by making a series expansion in the fugacity $e^{-\sigma}$. For example, strong coupling expansion was made for the $S^5$ partition functions of $U(N)$ gauge theories in a special unrefined limit \cite{Kim2013b, Kim2012}. The results give the $(2,0)$ superconformal indices of $SU(N)$-type, reproducing the BPS Kaluza-Klein spectrum of AdS${}_7 \times S^4$ supergravity in the large $N$ limit. Also, they take the form of the vacuum character of $\mathcal{W}_{A_{N-1}}$ algebra for finite $N$, leading to the $\mathcal{W}$ algebra conjecture \cite{Beem2015a}. We want to compute the $S^5$ partition functions for other gauge groups: $SO(2N)$, $SO(2N+1)$, $Sp(N)$. 

We follow the notations of \cite{Kim2013b,Kim2012}. The path integrals for the $S^5$ partition functions are reduced to the following matrix integrals
\begin{align}
	\label{eq:5-sphere}
	Z_{S^5} (\sigma, a,b,c,m)= \frac{1}{|W_G|}\int_{-\infty}^\infty \left[\prod_{i=1}^N d\lambda_i \right] e^{-\frac{2\pi^2 \, \text{Tr}\, \lambda^2}{\sigma (1+a)(1+b)(1+c)}} \cdot  Z_{\rm pert}^{(1)}Z_{\rm inst}^{(1)}  \cdot Z_{\rm pert}^{(2)}Z_{\rm inst}^{(2)}  \cdot Z_{\rm pert}^{(3)}Z_{\rm inst}^{(3)}. 
\end{align}
$N$ is the rank of 5d gauge group. $a$, $b$, $c$ are the $S^5$ squashing parameters $a, b, c$ satisfying $a+b+c=0$. $\sigma = \frac{g_5^2}{2\pi r_5}$ is the dimensionless coupling made of the $S^5$ radius $r_5$ and the Yang-Mills coupling $g_5$. We use the normalized trace $\text{Tr} (T^a T^b) = \delta^{ab}$ and set a long root $\vartheta$ of $G$ to satisfy $|\vartheta|^2 = 2$. The $S^5$ Yang-Mills action is written as $\frac{1}{4 g_5^2}\int \text{Tr}\,(F^{\mu\nu}F_{\mu\nu})$. The unit instanton action is $\frac{4\pi^2 }{\sigma}$ for all gauge groups. $Z_{\rm pert}$ are the indices for BPS bound states of perturbative W-bosons only. $Z_{\rm inst}$ are the instanton partition functions that we have studied so far. Omega-deformation parameters $\epsilon_{1}$, $\epsilon_{2}$ of $Z_{\rm pert}Z_{\rm inst}$ are identified with the squashing parameters $a$, $b$, $c$, such that $(\epsilon_{1},\epsilon_{2}) = (b-a,\,c-a)$, $(c-b,\,a-b)$, $(a-c,\,b-c)$ at each fixed point. See \cite{Kim2012} for the details.

Re-arranging the $S^5$ partition functions into strong coupling expressions is generally very difficult. Only the Abelian $(2,0)$ index \cite{Bhattacharya2008} was reproduced from the 5d $U(1)$ partition function on $S^5$ \cite{Kim2012}. Instead, we consider the special limit $m = \frac{1}{2}$ and $a, b, c \rightarrow 0$ in which the $S^5$ partition functions preserve 8 SUSYs. We can make the precise analysis in this limit as follows. The instanton partition functions are simplified such that the net instanton contributions come from only one fixed point. 
\begin{align}
	Z_\text{inst}^{(1)} \rightarrow 1\,,\quad 
	Z_\text{inst}^{(2)} \rightarrow 1\,,\quad
	Z_\text{inst}^{(3)} \rightarrow Z_\text{inst}|_{m=\epsilon_+}
\end{align}
if the limit $a, b, c \rightarrow 0$ is taken in a suitable order \cite{Kim2012}.
For all classical gauge groups, the instanton corrections are given as follows:
\begin{align}
	U(N),\ SO(2N),\ SO(2N+1):&\quad Z_\text{inst}^{(3)} = \text{PE} \left[ \frac{N\, e^{-4\pi^2/\sigma}}{1-e^{-4\pi^2/\sigma}}\right],\\
	Sp(N) \text{ with $\theta = 0$}:&\quad Z_\text{inst}^{(3)} = \text{PE} \left[ \frac{N\, e^{-8\pi^2/\sigma}}{1-e^{-8\pi^2/\sigma}} + \frac{e^{-4\pi^2/\sigma}}{1-e^{-8\pi^2/\sigma}}\right]\\
	Sp(N) \text{ with $\theta = \pi$}:&\quad Z_\text{inst}^{(3)} = \text{PE} \left[ \frac{N\, e^{-8\pi^2/\sigma}}{1-e^{-8\pi^2/\sigma}} + \frac{N\, e^{-8\pi^2/\sigma}}{1-e^{-16\pi^2/\sigma}}\right]
\end{align}
Multiplying the overall factors $e^{\frac{N \pi^2}{6\sigma}}$, $e^{(N-\frac{1}{2})\frac{\pi^2}{3\sigma}}$, $e^{(N-1)\frac{\pi^2}{3\sigma}}$ which correspond to the suitable couplings of the theories to the background curvatures \cite{Kim2013b},
the instanton contributions are written as Dedekind eta functions. We can easily make strong coupling expansions using the modular property of Dedekind eta function: $\eta(-\frac{1}{\tau}) = \sqrt{-i\tau} \eta(\tau)$. After all,
\begin{align}
	\label{eq:inst-s5}
		U(N),\ SO(2N),\ SO(2N+1): &\quad \left(\tfrac{2\pi}{\sigma}\right)^{N/2} e^{\frac{N \sigma}{24}}\  \text{PE} \left[ \frac{N\, e^{-\sigma}}{1-e^{-\sigma}}\right]\\
		Sp(N) \text{ with $\theta = 0$}: &\quad {2}^{-1/2} \left(\tfrac{4\pi}{\sigma}\right)^{N/2} e^{\frac{(N+1) \sigma}{48}}\  \text{PE} \left[ \frac{(N-1)\, e^{-\sigma/2}}{1-e^{-\sigma/2}} + \frac{e^{-\sigma}}{1-e^{-\sigma}}\right]\nn\\
		Sp(N) \text{ with $\theta = \pi$}: &\quad {2}^{-1/2}\left(\tfrac{4\pi}{\sigma}\right)^{N/2} e^{\frac{(N+1/2) \sigma}{48}}\  \text{PE} \left[ \frac{(N+1)\, e^{-\sigma/2}}{1-e^{-\sigma/2}} - \frac{e^{-\sigma/4}}{1-e^{-\sigma/4}}\right]\nn
\end{align}

We can also handle the matrix integrals \eqref{eq:5-sphere} over the perturbative determinants $e^{-2\pi^2 \, \text{Tr}\, \lambda^2/\sigma}\cdot Z_{\rm pert}^{(1)} Z_{\rm pert}^{(2)} Z_{\rm pert}^{(3)}$. For classical gauge groups $G$, the integrals become \cite{Kim2012}
\begin{align}
	\label{eq:pert-integral-s5}
	\left(\tfrac{\sigma}{2\pi}\right)^{N/2} e^{\frac{\sigma}{12}c_2 |G|} \prod_{\rho \in \Delta^+_G} 2\sinh{\left(\tfrac{
	\sigma \cdot \rho(\lambda)}{2}\right)} ,
\end{align}
where $\Delta^+_G$ is the set of positive roots of $G$. $c_2$ is the dual Coxeter number. $|G|$ is the dimension of the semi-simple part of $G$. $\rho$ is the Weyl vector. The final results are products of \eqref{eq:inst-s5} and \eqref{eq:pert-integral-s5}. 

We summarize the exact $S^5$ partition functions for all classical gauge groups as follows:
\begin{align}
	\label{eq:5-sphere-sun}
	Z_{S^5}^{SU(N)} &\equiv Z_{S^5}^{U(N)}/Z_{S^5}^{U(1)} =  e^{\frac{\sigma}{6}c_2 |G| + \frac{\sigma (N-1)}{24}} \cdot \text{PE} \left[  \frac{\sum_{m=2}^N e^{-m\sigma}  }{1-e^{-\sigma}} \right]\\
	\label{eq:5-sphere-so2n}
	Z_{S^5}^{SO(2N)} &= e^{\frac{\sigma}{6}c_2 |G| + \frac{\sigma N}{24}}\cdot \text{PE} \left[ \frac{e^{-N\sigma} + \sum_{m=1}^{N-1} e^{-2m\sigma} }{1-e^{-\sigma}} \right]\\
	\label{eq:5-sphere-so2np1}
	Z_{S^5}^{SO(2N+1)} &= e^{\frac{\sigma}{6}c_2 |G| + \frac{\sigma N}{24}}\cdot \text{PE} \left[ \frac{-e^{-\sigma/2}+ e^{-(N+\frac{1}{2})\sigma} + \sum_{m=1}^N e^{-(2m-1)\sigma}   }{1-e^{-\sigma}} \right]\\
	\label{eq:5-sphere-spn}
	Z_{S^5}^\text{Sp(N)$|_{\theta=0}$} &= 2^{(N-1)/2} \, e^{\frac{\sigma}{6}c_2 |G| + \frac{ (N+1)\sigma}{48}} \cdot \text{PE} \left[ \frac{-e^{-\sigma/2}-e^{-\frac{N+1}{2}\sigma} -e^{-\frac{N+2}{2}\sigma}+ \sum_{m=0}^N \ (e^{-\frac{2m+1}{2}\sigma} + e^{-\frac{2m+2}{2}\sigma})  }{1-e^{-\sigma}} \right]\\
	Z_{S^5}^\text{Sp(N)$|_{\theta=\pi}$} &= 2^{(N-1)/2} \, e^{\frac{\sigma}{6}c_2 |G| + \frac{(N+1/2) \sigma}{48}} \cdot \text{PE} \left[ \frac{- e^{-\sigma/4}-e^{-\frac{N+1}{2}\sigma}  + \sum_{m=0}^N e^{-\frac{2m+1}{2}\sigma}   }{1-e^{-\sigma/2}}\right] 
\end{align}
We interpret them as the indices of 6d $(2,0)$ theories. Recall that $(2,0)$ theories follow the ADE classification. The gauge theories  with simply-laced groups uplift to  $(2,0)$ theories on $S^1$. The gauge theories with non-simply-laced  groups are proposed to uplift to circle compactified $(2,0)$ theories with outer automorphism twists. The $S^5$ partition functions of $SU(N)$ and $SO(2N)$ gauge theories correspond to the $(2,0)$ superconformal indices of $SU(N)$-type and $SO(2N)$-type. Note that the vacuum Casimir energies scale as $N^3$ as expected from dual gravity \cite{Klebanov1996,Awad2001}. The $SU(N)$ superconformal indices were already obtained in \cite{Kim2012}, which agree with the vacuum character of $\mathcal{W}_{A_{N-1}}$ algebra. Our results show that the $SO(2N)$ superconformal indices also agree with the vacuum character of the $\mathcal{W}_{D_N}$ algebra, providing the evidence that the $\mathcal{W}$ algebra conjecture made in \cite{Beem2015a} holds for $SO(2N)$ theories.

Let us turn to the $SO(2N+1)$ or $Sp(N)$ partition functions.  These non-simply-laced gauge theories are obtained from ADE-type $(2,0)$ theories on $S^1$ with outer automorphism twists \cite{Witten2009,Tachikawa2011b}: 
\begin{enumerate}
  \setlength{\itemsep}{0pt}
	\item $SU(2N)$-type $(2,0)$ theories with $\mathbb{Z}_2$ outer automorphism give 5d $SO(2N+1)$ SYMs.
	\item $SO(2N+2)$-type $(2,0)$ theories with $\mathbb{Z}_2$ outer automorphism give 5d $Sp(N)$ SYMs with $\theta=0$.
	\item $SU(2N+1)$-type $(2,0)$ theories with $\mathbb{Z}_2$ outer automorphism give 5d $Sp(N)$ SYMs with $\theta=\pi$.
\end{enumerate}
We expect that the $S^5$ partition functions for non-simply-laced gauge theories to be the twisted indices of corresponding $(2,0)$ theories. To propose them as the new indices for $(2,0)$ theories with twists, we remove by hand the overall numerical factor $2^{(N-1)/2}$ in \eqref{eq:5-sphere-spn} to adjust the ground state contribution to the index to be $1$. This may have to do with a suitable choice of path integral measures for the 6d theories with outer automorphism twists. However, we have no good explanations at the moment. These indices are compatible with the twisted compactification realizations of non-simply-laced gauge theories, in that they have smooth large $N$ limits. In the large $N$ limit, these indices are given as 
\begin{align}
	\label{eq:superconformal-index-soodd}
	Z_{S^5}^{SO(2\infty+1)} &= \text{PE} \left[ -\frac{e^{-\sigma/2}}{1-e^{-\sigma}} + \frac{e^{-\sigma}}{(1-e^{-\sigma})(1-e^{-2\sigma})} \right]\\
	\label{eq:superconformal-index-sp}
	Z_{S^5}^{Sp(\infty)_{\theta=0}} &= \text{PE} \left[-\frac{e^{-\sigma/2}}{1-e^{-\sigma}}+  \frac{e^{-\sigma/2}}{(1-e^{-\sigma/2})(1-e^{-\sigma})} \right]\\
	Z_{S^5}^{Sp(\infty)_{\theta=\pi}} &= \text{PE} \left[-\frac{e^{-\sigma/4}}{1-e^{-\sigma/2}}+  \frac{e^{-\sigma/2}}{(1-e^{-\sigma/2})(1-e^{-\sigma})} \right].
\end{align}
These results may shed more lights on what one precisely means by outer automorphism twists of 6d $(2,0)$ theories.

\section{Conclusion}
\label{sec:conclusion}
In this paper, we studied the instanton partition functions for 5d $\mathcal{N}=2$ SYM with classical gauge groups. Our approach was to utilize the ADHM quantum mechanics of the D0-D4-O4 systems. The instanton partition functions were obtained by computing their Witten indices and extracting out the extra states' index contribution. The extra factor captured the BPS spectrum of D0-branes in pure orientifold backgrounds. We also studied S-dualities of various O4-planes on $S^1$.

The instanton partition functions respect 6d S-dualities which are uplifts of 4d S-dualities. Using them, we also computed the $S^5$ partition functions which correspond to the 6d superconformal indices. In the special limit, our final results showed that the superconformal indices of $SO(2N)$-type $(2,0)$ theories agree with the vacuum character of $\mathcal{W}_{D_N}$ algebra. The instanton partition functions for non-simply-laced $SO(2N+1)$, $Sp(N)|_{\theta=0}$, $Sp(N)|_{\theta=\pi}$ groups also produced interesting results. They are proposed to be the indices of $(2,0)$ theories with outer automorphism twists. 

It would be important to understand S-dualities of the instanton partition functions on $\mathbb{R}^4 \times S^1$ in full generality. This will be useful to study the spectrum of $(2,0)$ theories via the superconformal indices. More generally, it would allow us to have better understanding on the high temperature behavior of $(2,0)$ theories.

\vspace{\baselineskip}
\noindent{\bf\large Acknowledgements}

\noindent
We thank Chiung Hwang, Sung-Soo Kim, Futoshi Yagi, and especially Kimyeong Lee, Piljin Yi  for helpful discussions. The work of YH and SK is supported in part by the National Research Foundation of Korea (NRF) Grant NRF-2015R1A2A2A01003124.

\providecommand{\href}[2]{#2}\begingroup\raggedright\endgroup

\end{document}